\documentclass{emulateapj}
\usepackage{epstopdf}
\usepackage{graphicx}

\shorttitle{HI in Filaments and Tendrils}
\shortauthors{Odekon et al.}
\begin{document}

\title{The Effect of Filaments and Tendrils on the HI Content of Galaxies}
\author{Mary Crone Odekon}
\affil{Department of Physics, Skidmore College, Saratoga Springs, NY 12866, USA; mcrone@skidmore.edu}
\author{Gregory Hallenbeck}
\affil{Washington \& Jefferson College, Department of Computing and Information Studies, 60 S Lincoln Street, Washington PA, 15301, USA; ghallenbeck@washjeff.edu} 
\author{Martha P. Haynes}
\affil{Cornell Center for Astrophysics and Planetary Science, Cornell University, Ithaca, NY 14853, USA; haynes@astro.cornell.edu}
\author{Rebecca A. Koopmann}
\affil{Department of Physics and Astronomy, Union College, Schenectady, NY 12308, USA; koopmanr@union.edu}
\author{An Phi}
\affil{Department of Physics, Skidmore College, Saratoga Springs, NY 12866, USA; aphi@skidmore.edu}
\author{Pierre-Francois Wolfe}
\affil{Lincoln Laboratory, Massachusetts Institute of Technology, Lexington, MA 02421, USA; Pierre-Francois.Wolfe@ll.mit.edu}

\begin{abstract}
 We use the ALFALFA HI survey to examine whether the cold gas reservoirs of galaxies are inhibited or enhanced in large-scale filaments.  Our sample includes 9947 late-type galaxies with HI detections, and 4236 late-type galaxies with well-determined HI detection limits that we incorporate using survival analysis statistics.
 We find that, even at fixed local density and stellar mass, and with group galaxies removed, the HI deficiency of galaxies in the stellar mass range $8.5<$log(M/M$_\odot)<10.5$ decreases with distance from the filament spine, suggesting that galaxies are cut off from their supply of cold gas in this environment.   
 We also find that, at fixed local density and stellar mass, the galaxies that are the most gas-rich are those in small, correlated ``tendril" structures within voids: although galaxies in tendrils are in significantly denser environments, on average, than galaxies in voids, they are not redder or more HI deficient. This stands in contrast to the fact that galaxies in tendrils are more massive than those in voids, suggesting a more advanced stage of evolution.  Finally, at fixed stellar mass \textit{and color}, galaxies closer to the filament spine, or in high density environments, are more deficient in HI.  This fits a picture where, as galaxies enter denser regions, they first lose HI gas and then redden as star formation is reduced.

\end{abstract}


\section{Introduction} 

The large-scale structure of the universe presents physically distinct environments for galaxy evolution.  In virialized groups and clusters, galaxies orbit around a common center of mass, sometimes at high speed and through hot gas.  
Filaments channel galaxies toward clusters, while bubble-like voids harbor
galaxies that have not yet been gravitationally pulled into denser regions. 
Of particular importance is the fact that, while many studies have found a dependence of galaxy properties on local density (e.g., Dressler 1980; Postman \& Geller 1984; Balough et al. 2004; Tanaka et al. 2004; Blanton et al. 2005; Mouhcine et al. 2007; Park et al. 2007; Grutzbauch et al. 2011; Brough et al. 2013; Rawle et al. 2013; Lai et al. 2016; Odekon et al. 2016), local density is not a perfect proxy for large scale structure.  
Indeed, the ranges in local density for groups, filaments, and voids overlap considerably (Libeskind et al. 2017). Given the importance of understanding these environments to models of galaxy evolution, there are a growing number of methods for quantifying large-scale environments, tailored to the information available in particular simulations and observations. Libeskind et al. (2017) provide a review and a comparison of different methods.

We might therefore expect cold gas accretion to depend on large-scale structure as well as local density.  Cold gas accretion from the cosmic web is likely to be a dominant factor in the growth of galaxies (e.g. Conselice et al. 2013; see Sanchez Almeida 2014 for a review).  Observations that may indicate cold gas accretion include streams of HI gas surrounding M31 (Thilker et al. 2004), HI in red early-type dwarf galaxies on the outskirts of the Virgo cluster (Hallenbeck et al. 2012, but see Hallenbeck et al. 2017 for follow-up observations that suggest the HI might not be recently accreted), and an HI envelope surrounding a small extended grouping of galaxies within a void (Beygu et al. 2013).  At the same time, cold gas is expected to be removed or cut off from galaxies through a variety of physical mechanisms that depend on large-scale environment.  In clusters, interactions with the hot intergalactic medium should remove gas through ram pressure stripping (e.g. Tonneson \& Bryan 2009) and the Kelvin-Helmholtz instability (e.g. Cen \& Riquelme 2008).  The gas cooling time in clusters may be too long to allow cool gas flows (Dekel \& Birnboim 2006), further starving galaxies of new cold gas.  Interactions with other galaxies should deplete cold gas reservoirs through major mergers, low-speed tidal encounters in small groups, and repeated fast-speed encounters (Moore et al. 1995).  

A particular question regarding the physical effect of large-scale filaments is whether they primarily act as conduits of cold gas that can replenish galaxies, or alternatively, as regions where galaxies are cut off from cold gas. Simulations highlight several different physical processes that may be relevant. Bahe et al. (2013), for example, use high-resolution hydrodynamical simulations to find that ram pressure is enhanced not only in clusters, but along the nearby parts of connecting filaments. Aragon-Calvo et al. (2016) describe the cessation of cold gas accretion through dynamical ``cosmic web detachment."  In this picture, galaxies accrete cold gas from the cosmic web until they enter regions where velocity streams cross, including regions near the central spine of large filaments.

In support of the picture where filaments replenish gas are the results of Kleiner et al. (2017) on the HI content of large galaxies.  They
use spectral stacking of galaxies in the HI Parkes All-Sky Survey to show that high-mass galaxies within 0.7 Mpc of filament spines  have a higher HI fraction than a control sample of galaxies more than 5 Mpc from filament spines.  The HI fraction is evaluated for each sample in three bins of projected density to control for the already-known dependence of galaxy properties on density, and both samples are limited in local density to remove groups. These results are specifically for galaxies with log(M$_{*}$/M$_\odot)>11$; for galaxies with smaller stellar masses, they report no difference between filament galaxies and the control sample.  

On the other hand, most results support a picture where galaxies in filaments are cut off from their supply of gas relative to galaxies in voids, leading to the quenching of star formation.  Alpaslan et al. (2016) find that isolated spiral galaxies in the Galaxy and Mass Assembly (GAMA) survey have higher stellar masses and lower specific star formation rates near the central spine of filaments than near the periphery. Similarly, in the COSMOS2015 catalog (Laigle et al. 2017), more massive galaxies are statistically closer to filaments, and at fixed stellar mass, passive (red) galaxies are closer to the filament when the contribution of nodes (groups) and the effects of local density are removed. 
Kuutma et al. (2017) find that at fixed density, high-mass galaxies ($M_r<-20$ mag) within filaments have reduced star formation rates (but not higher masses) and are more likely to have early-type morphologies.  Poudel et al. (2017) find that at fixed group mass and environmental density, groups in filaments are more luminous, and central galaxies in filaments have redder colors, higher masses, and lower specific star formation rates.

It is worth emphasizing that at fixed galaxy type and mass, it is much harder to see the effect of quenching.  For example, Moorman et al. (2016) find no difference in star formation efficiency (star formation relative to HI mass) or in specific star formation rate at constant stellar mass for void galaxies relative to non-void galaxies once the sample is limited to galaxies detected in HI (which tend to be overwhelmingly late-type galaxies). 

Studies of additional galaxy properties also show a dependence on large-scale structure. For example, Chen et al. (2016) find that at the same local density, galaxies closer to filaments are more massive, in agreement with Alpaslan et al. (2016) and Laigle et al. (2017). Guo et al. (2015) find that the satellite luminosity function for isolated central galaxies in filaments is shifted higher than it is for isolated central galaxies outside of filaments, so that satellites are more abundant in filaments.  Tempel et al. (2015) examine the alignments between filament axes and the positions of satellites and central galaxies and find they are correlated, especially for bright and red galaxies. Zhang et al. (2015) find a weak alignment between spiral spin axes and the large-scale tidal field.

To summarize, both theoretical and observational results suggest that large-scale structure has an important affect on galaxy evolution beyond the known trends with local density. On the specific problem of quenching, recent studies indicate that galaxies close to filament spines have lower specific star formation rates.  However, observations of the HI gas that could fuel star formation are more ambiguous: for large galaxies, the HI content seems to be \textit{enhanced} near filaments, while the star formation efficiency in voids is not distinguishable from that in denser environments.  

In this paper, we use HI observations from the ALFALFA survey (Giovanelli et al. 2005) to examine the cold gas content of galaxies as a function of large scale structure. To identify structures, we use a modified minimal spanning tree method (following Alpaslan et al. 2014) that we find to be well-suited to our observational data.

The paper is organized as follows: we describe the calculation of galaxy properties in Section 2, the determination of large-scale environment and local density in Section 3, and results in Section 4.  We summarize and discuss our conclusions in Section 5. Throughout this paper we use the cosmological parameters $\Omega_{m}=0.3$, $\Omega_{\Lambda}=0.7$, and $h=0.7$.

\section{Data}

\subsection{HI and optical properties}

Our procedure for calculating HI and optical properties is very similar to that in Odekon et al. (2016) in the context of quantifying the amount of HI in groups and clusters. Our HI data are from the $\alpha.70$ catalog, the most recent data release from the ALFALFA survey (Giovanelli et al. 2005), which covers 70\% of the total survey area of 7000 deg$^{2}$ with an rms noise $\sim 2.2$ mJy and beam size $\sim 3.6$ arcminutes.  
Because we use the DR7 Yang groups catalog (see Yang et al. 2007) to define filaments, we select a smaller region of the $\alpha.70$ catalog that is embedded within the volume covered by the groups catalog.  Figure 1 shows this region of sky.  The range in redshift is 0.015 to 0.05.   

\begin{figure*}
	\includegraphics*[width=500pt]{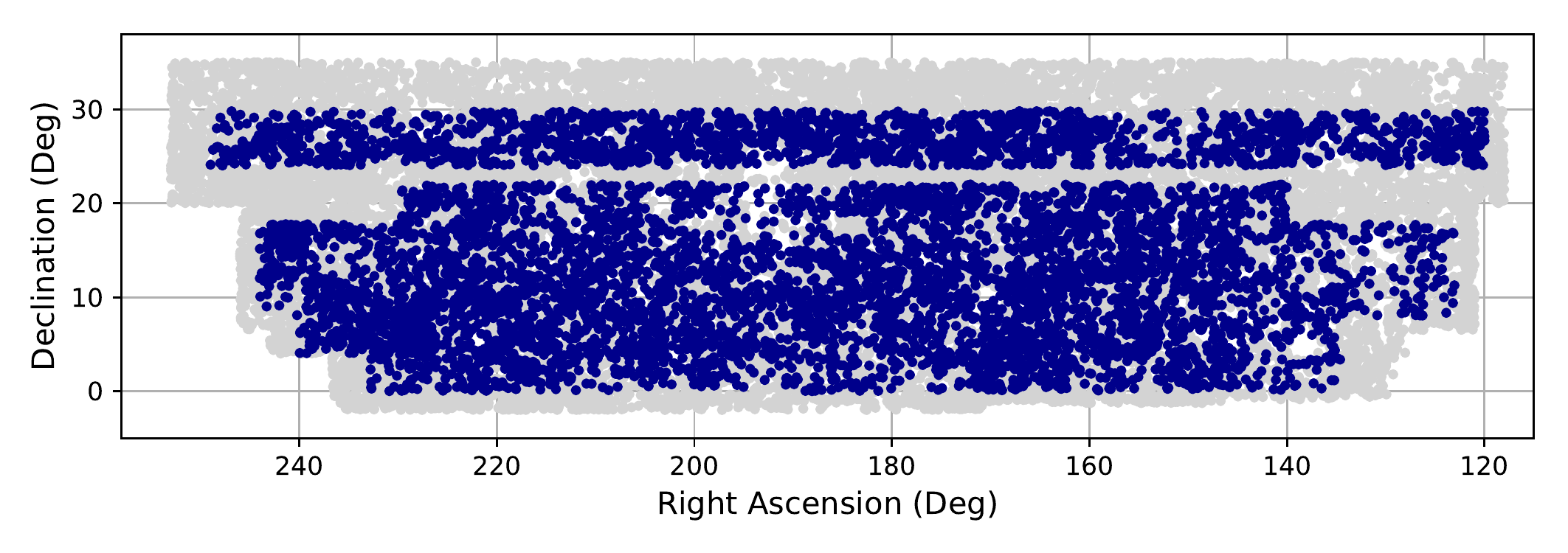}
	\setlength{\abovecaptionskip}{-5pt} 
	\setlength{\belowcaptionskip}{0pt} 
	\caption{Sky map of the sample from the ALFALFA 70\% survey (blue) and the surrounding region (gray) used to determine galaxy environment.}
\end{figure*}

For galaxy colors and stellar masses, we use optical data from the Sloan Digital Sky Survey (SDSS; Albareti et al. 2016). Following the recommendations on the SDSS web site, we use SDSS cmodel mags for galaxy magnitudes and model mags for colors.  We use galactic extinction corrections from Schlafly \& Finkbeiner (2011) via the map tool at the NASA/IPAC Infrared Science Archive, and internal extinction corrections from Shao et al. (2007).  
We estimate stellar masses using Equation (7) in Taylor et al. (2011): $\log M_{\star}/L_{i}=-0.68+0.70(g-i)$; we find good overall agreement between these masses and those from SED fitting models for a subset of galaxies in our sample where the SED fitting models are available on the SDSS web site.\footnote[1]{www.sdss.org/dr12/spectro/galaxy/}

Identifications between ALFALFA detections and SDSS detections are verified by inspection before inclusion in the matched ALFALFA catalog to avoid spurious objects (like individual star forming regions within a galaxy) that the SDSS pipeline may identify as a separate galaxy.  The total number of ALFALFA-detected galaxies within our subset of the $\alpha.70$ catalog volume and with spectrosopic SDSS matches is 11,375.

We describe galaxy HI content through the HI deficiency. First, HI masses are calculated using $M_{HI}/M_{\sun}=2.356\times 10^{5}D_{Mpc}^{2}S_{int}$, where $S_{int}$ is the integrated ALFALFA flux in Jansky km s$^{-1}$ (e.g Giovanelli et al. 2005).  
HI deficiency is the logarithmic difference between the observed HI mass and the HI mass expected based on galaxy size:  $HIDef=\log M_{HI}^{exp}-\log M_{HI}^{obs}$ (Haynes et al. 1984).  We use the calibration of Toribio et al. (2011), who find that $M_{HI}$ correlates especially well with galaxy diameter according to the relationship $\log (M_{HI}/M_{\sun})=8.72+1.25\log D_{25}$. 
As in Toribio et al. (2011), the r-band diameter $D_{25}$ in kiloparsecs is from the size $isoA_{r}$ in the SDSS DR7, the galaxy inclination from the axial ratio $expAB$, and the number of kiloparsecs per arcsecond at the distance of the galaxy $adist$ according to $\log D_{25} = \log(isoA_{r}0.39 \arcsec adist)+0.35\log(expAB)$. 
Distances are estimated from the Hubble flow, using the average redshift of the Yang group, or the SDSS redshift of the individual galaxy if it is not a member of a Yang group. 

To select late-type galaxies, we consider both the galaxy color-magnitude diagram and classifications in the Galaxy Zoo (Lintott et al. 2008).  First, to identify members of the blue cloud, we fit the color-magnitude diagram to two elliptical Gaussian distributions and exclude any galaxy identified as more than 1\% likely to be part of the elliptical distribution corresponding to the red sequence. (We used a similar procedure in Odekon et al. 2016; see Figure 2 of that paper.)  Using the color-magnitude diagram allows us to include galaxies that are clearly members of the blue cloud, but that do not have a distinct spiral structure according to the Galaxy Zoo. The number of HI-detected galaxies in our sample thus identified is 9820.  This method, however, excludes many large spirals at the bright, red end of the blue cloud.  We augment our sample with galaxies satisfying the Galaxy Zoo ``clean" spiral criteria (Lintott et al. 2008), where at least 80\% of users classify a galaxy as right-hand spiral or left-hand spiral. While most of these are already included in the blue cloud sample, this provides an additional 127 spirals, most of which populate the massive, red end of the blue cloud, for a total of 9947 late-type galaxies with HI detections.

\subsection{HI detection limits}

In addition to statistics based on the galaxies detected by the ALFALFA, we use survival analysis techniques that take into account detection limits for late-type galaxies not detected as HI sources. Specifically, we use the cenreg function within the NADA package in the R statistics environment to run censored regressions using maximum likelihood estimation (Helsel 2012).

We identify late-type galaxies not detected in HI from spectroscopically-observed SDSS galaxies within the ALFALFA survey region and not closer than 30" from a galaxy detected in HI, in order to exclude galaxies that might be confused with another source.  In addition, the ALFALFA spectra must be free from any features that could hide a possible detection or interfere with estimating the rms noise for that region, which we use to estimate the HI limiting flux. These features include radio frequency interference, baseline problems, and confusion with other nearby galaxies. To identify clean spectra, we extract spectra from ALFALFA grids following the procedure described by Fabello et al. (2011). We integrate the grid over a region 4x4 arcmin$^{2}$ centered on the optical position of the galaxy and over all velocities. Inspecting a random subsample of 600 spectra, we find that problematic spectra are eliminated by selecting only those with a skew in flux per channel between -0.05 and 0.15, and an rms in flux per channel less than 3.0 mJy.  This selection process yields 4236 late-type galaxies within our sample volume with clean spectra and no detection in ALFALFA.

To calculate the rms noise needed for estimating the limiting flux, we use the method described in Odekon et al. (2016).  The rms for each polarization is determined outside of the estimated channel range of the galaxy and eliminating low weight regions $(<50\%)$. 
As discussed in Giovanelli et al. 2005, ALFALFA grids are constructed from individual drift scans for each of seven beams and two polarizations with spectral sampling at one-second intervals. Sometimes one or more of the 14 spectra may not have been available due to hardware failures. To accommodate this, each spatial pixel in the grid is assigned a weight based on how many individual spectra contributed to the final grid at that position. Further weighting in the spectral domain accounts for the flagging of individual channels for radio frequency interference.

The channel range of the galaxy is estimated using the Tully-Fisher relation and the SDSS i-band magnitude following Giovanelli et al. (1997), adopting $\gamma=1.0$ in the internal extinction correction $\Delta m = - \gamma \log(a/b)$ (their equation 8), which is appropriate for all but the brightest galaxies and overcorrects for fainter galaxies (see Figure 7c of Giovanelli et al. 1995); this ensures the rms is measured outside any possible galaxy signal. We fit the baseline with a first-order polynomial and calculate the rms about this fit, using the Arecibo IDL program robfit\_poly.pro by Phil Perrilat. Our process is conservative in the sense that we exclude a large spectral region around each galaxy just in case there is any signal from those that are not detected in HI; comparing these results with those from a subset where we use the less conservative estimate  $\gamma=0.5$, we see no significant differences.

The limiting flux for galaxies not detected by ALFALFA can then be estimated as 
$$S_{21,lim}= \sqrt{W_{50}\ 20}\  \sigma_{rms}\ S/N_{lim}$$ 
where $W_{50}$ is the expected 21-cm line width, the signal to noise $S/N_{lim}$ is set at 6.5, and the rms noise $\sigma_{rms}$ is measured directly for the location of each optically-detected galaxy not detected in HI (Gavazzi et al. 2013).  The signal to noise calculation for the ALFALFA pipeline is described in Saintonge (2007), and the statistical properties of the ALFALFA dataset are described in Haynes et al. (2011).

We estimate W$_{50}$ following Springob et al. (2007): 
$$M_{I}=-7.85(\log W-2.5)-20.85+\log h.$$ 
The width $W$ must be corrected for inclination and broadening to give the expected observed width $W_{50}$.  
To correct for inclination, we multiply the width by $\sin i$, where the inclination angle $i$ is determined from the axial ratio according to 
$\cos^{2}i=((b/a)^{2}-q^{2})/(1-q^{2})$  where $b/a$ is the axial ratio from the SDSS and $q=0.20$ (see Springob et al. 2007). When $b/a < q$, $i$ is set to $90\degr$.  Finally, we add 6.5 km/s to account for broadening by internal motions, multiply by $1+z$ to account for broadening by redshift, and add 10 km/s to account for instrumental broadening, as described in Springob et al. (2007). 

\section{Large-scale structure and local density}

To characterize the large-scale and local environment for each galaxy, we use a separate, magnitude-limited data set from the SDSS that surrounds the main data sample described above.  We select this region to lie within that of the Yang group catalog (Yang et al. 2007) since we use the positions of Yang groups to determine filaments. The boundaries of the region are trimmed to a simple shape to minimize the identification of spurious structures. Figure 1 shows the sky positions of both the ALFALFA sample and the environment data set.  The environment data set extends out to z=0.055, slightly past the ALFALFA sample, which ends at z=0.050.  This provides a buffer zone to help characterize the environment properly for galaxies near the edge of the main sample.

To create a magnitude-limited sample, we include only galaxies with an absolute r-band magnitude brighter than -19.16, corresponding to an apparent r-band magnitude of 17.7 at the far end of our sample volume at z=0.055. (See http://www.sdss.org/dr12/algorithms/magnitudes/ for a discussion of completeness limits for SDSS galaxies.)  This process yields 40,103 galaxies for the environment data set.

We define large-scale structure using our own implementation of the technique developed by Alpaslan et al. (2014) for use with the Galaxy and Mass Assembly Large-Scale Structure Catalogue (GLSSC).  We find that, although our sample volume has a very different shape from the GLSSC, covering a much larger part of the sky but not extending nearly as far in redshift, this technique is effective and straightforward to implement. 

First, the centers of Yang DR7 groups are used as nodes in a minimal spanning tree (MST) to define filaments.  Only groups closer than a separation distance \textit{b} from other groups are included as filament nodes.  We use groups that have a minimum mass of $2\times10^{12}$ M$_{\odot}$ according to the luminosity-based halo mass estimate in the Yang catalog, a population that is complete throughout our sample volume.  This population is  similar to the group population used for the GLSSC (although they define their groups in a different way; see figure 2 in Alpaslan et al. 2014), so we can use the same linking length they do, $b= 5.75h^{-1}$ Mpc, to define the same types of filaments. The number of groups in our data set is 4868, about 80\% of their 6000. This is consistent with the relative sizes of the sample volumes: ours is approximately $(157 h^{-1} Mpc)^{3}$, about 80\% of their $(167 h^{-1} Mpc)^{3}$. 

After removing galaxies closer than $4.12h^{-1}$ Mpc to the filaments, Alpaslan et al. (2014) found that there are still smaller, filamentary structures they call ``tendrils'' embedded within the voids.  To identify these structures, they created an additional MST on the remaining galaxies (not the group centers this time) trimmed with a linking length of $4.13 h^{-1}$ Mpc.  We use a smaller linking length of $3.0 h^{-1}$ Mpc, to account for the our fainter limiting magnitude (-19.16 instead of -19.77), and correspondingly smaller average distance between galaxies. Galaxies remaining after tendrils are removed are considered void galaxies.

The large-scale environment for each galaxy can thus be categorized as group, filament, tendril, or void. In addition, the perpendicular distance of each galaxy to the central spine of the nearest filament can be used as a continuous variable to examine the effect of the filament environment. The perpendicular distance metric was found to be useful in examining the effect of filaments on the specific star formation of isolated late-type galaxies by Alpaslan et al. (2016). 

To determine the environment category for each galaxy in the main sample of HI detections and detection limits, we match each galaxy in the main sample to the closest galaxy in the environment data set within $cz \pm 500$ km/s and assign the environment of that galaxy. Following this procedure, we find the following breakdown for the number of galaxies in our main sample with HI detections and well-determined detection limits: 1093 detections and 795 limits in groups, 4528 detections and 1898 limits in filaments, 3425 detections and 1185 limits in tendrils, and 901 detections and 358 limits in voids. To determine the distance to the nearest filament, we take each galaxy in the main sample and directly calculate the three-dimensional perpendicular distance to the nearest filament spine.  Figure 2 illustrate the structures identified within the northern slice of our main data set (between 22 and 25 degrees in declination).   This is the region of the CfA2 Great Wall, which appears as arcs on either side of the Coma cluster, and a large spur at higher redshift and lower right ascension. Note that tendrils do not look like small groups in the sense that they are not, in general, elongated along line of sight.

\begin{figure*}
	\includegraphics*[width=500pt]{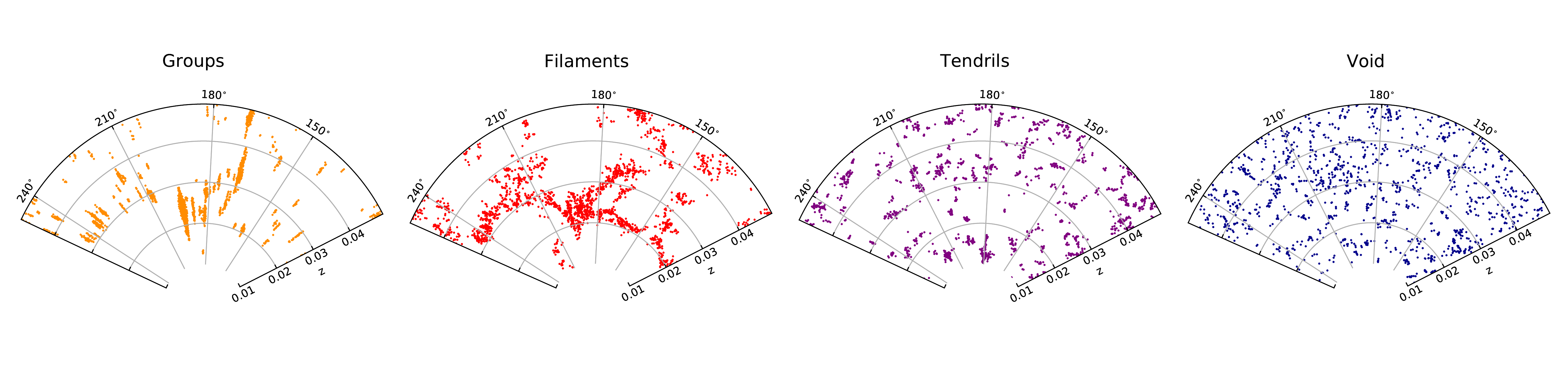}
	\setlength{\abovecaptionskip}{0pt} 
	\caption{Distribution in right ascension and redshift of galaxies in the northern slice with declination between 22 deg and 25 deg, showing large-scale structure as categorized by membership in groups, filaments, tendrils, or void.  Coma is the large group at R.A. $\sim 200\deg$, and the CfA2 Great wall extends across the region near z=0.02 to z=0.035.}
	\vspace{-0mm}
\end{figure*}

We use the third nearest-neighbor density to compare the effects of large-scale structure with those of local density.  Our procedure is very similar to that of Jones et al (2016).  For each galaxy in the main sample, we find the distance $r$ on the sky to the third nearest neighbor in the environment data set that is within $cz \pm 500$ km/s.  Any galaxy in the environment data set closer than 5 arcseconds with $cz \pm 70$ km/s is discarded, as this nearly always corresponds to the galaxy from the main sample that is under consideration, rather than a separate nearby galaxy.  The third nearest-neighbor density is $\Sigma_{3} = 3 \pi^{-1} r^{-2}$.  

\begin{figure*}
	\includegraphics*[width=500pt]{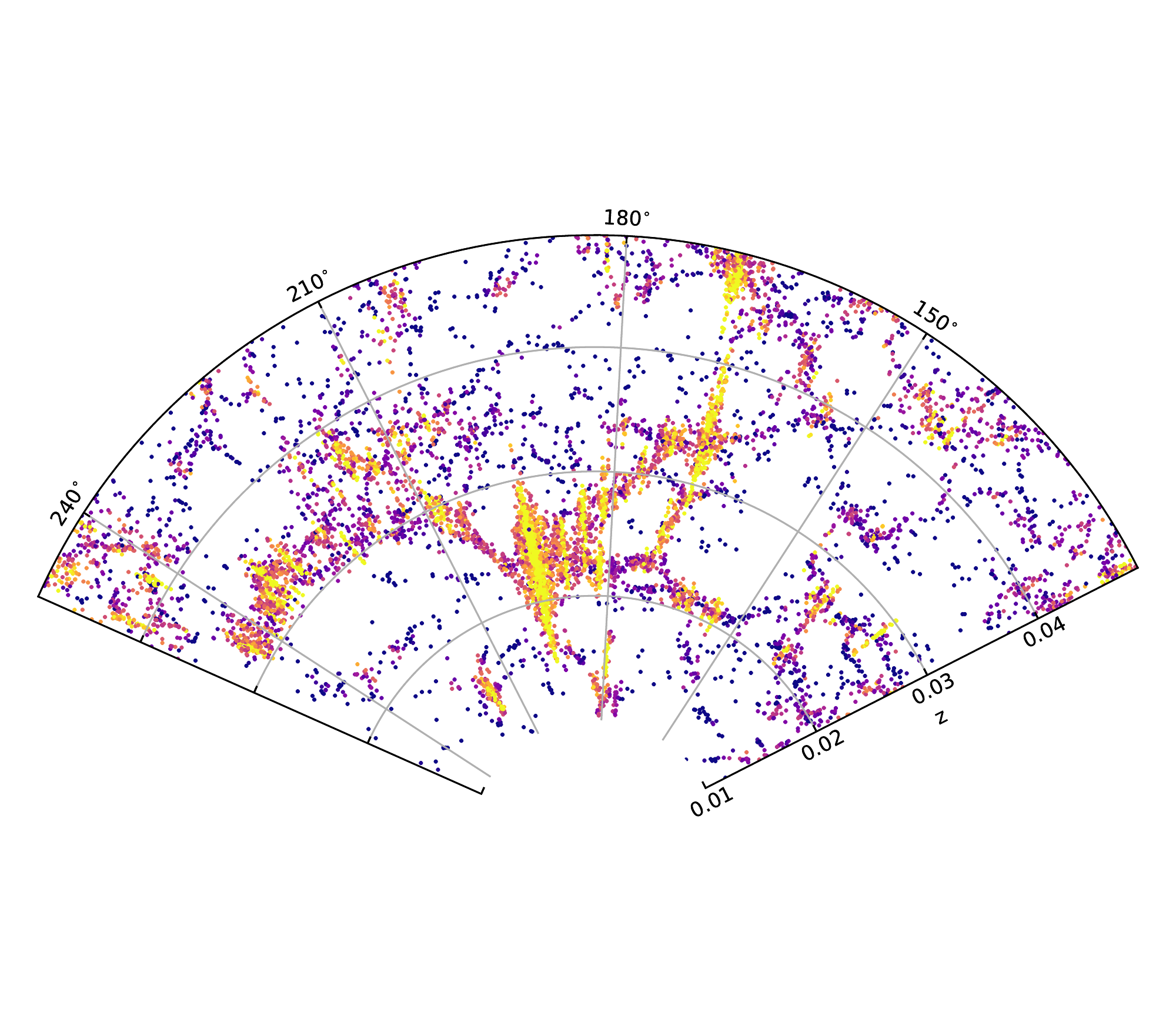}
	\setlength{\abovecaptionskip}{-50mm}  
	\caption{Like Figure 2, but including all galaxies in the slice, color-coded by nearest-neighbor density.}
	\vspace{-20mm}
\end{figure*}

Figure 3 shows the galaxies in the northern slice color-coded by local density, and Figures 4 and 5 compare the large-scale structure metrics with local density.  The average local density increases systematically from void to tendril to filaments to groups (Figure 4), although with a significant dispersion in values for each category. (Even within a specific system, the local density varies considerably among galaxies; the average value of Log $\Sigma_3$ for galaxies in the Coma cluster, for example is 0.62 with a standard deviation of 0.41.) 

The logarithm of the distance to the filament spine, Log D$_{fil}$, is roughly inversely proportional to Log $\Sigma_3$ (Figure 5).  An exception to the approximately inverse relationship is the concentration of galaxies at moderate density but Log D$_{fil} \sim -1$; these galaxies are in groups that are relatively small but that define the filament spine.

\begin{figure}
	\epsscale{1.0}
	\includegraphics*[width=250pt]{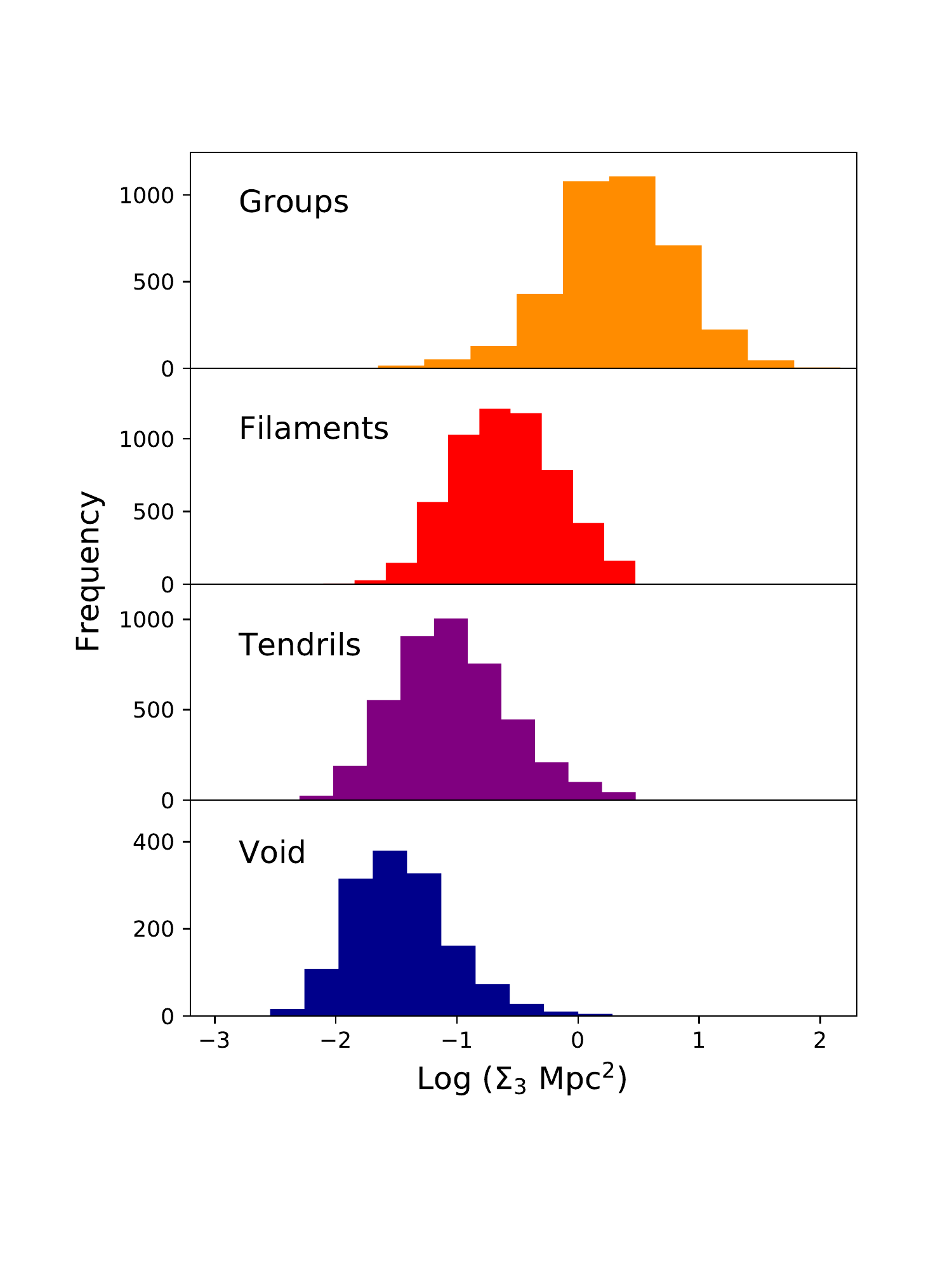}
	\setlength{\abovecaptionskip}{-80pt} 
	\caption{Histograms of local density for each category of large-scale structure. The local density increases systematically from void to group galaxies.}  
	\vspace{-10mm}
\end{figure}

\begin{figure*}
	\epsscale{1.0}
	\includegraphics*[width=500pt]{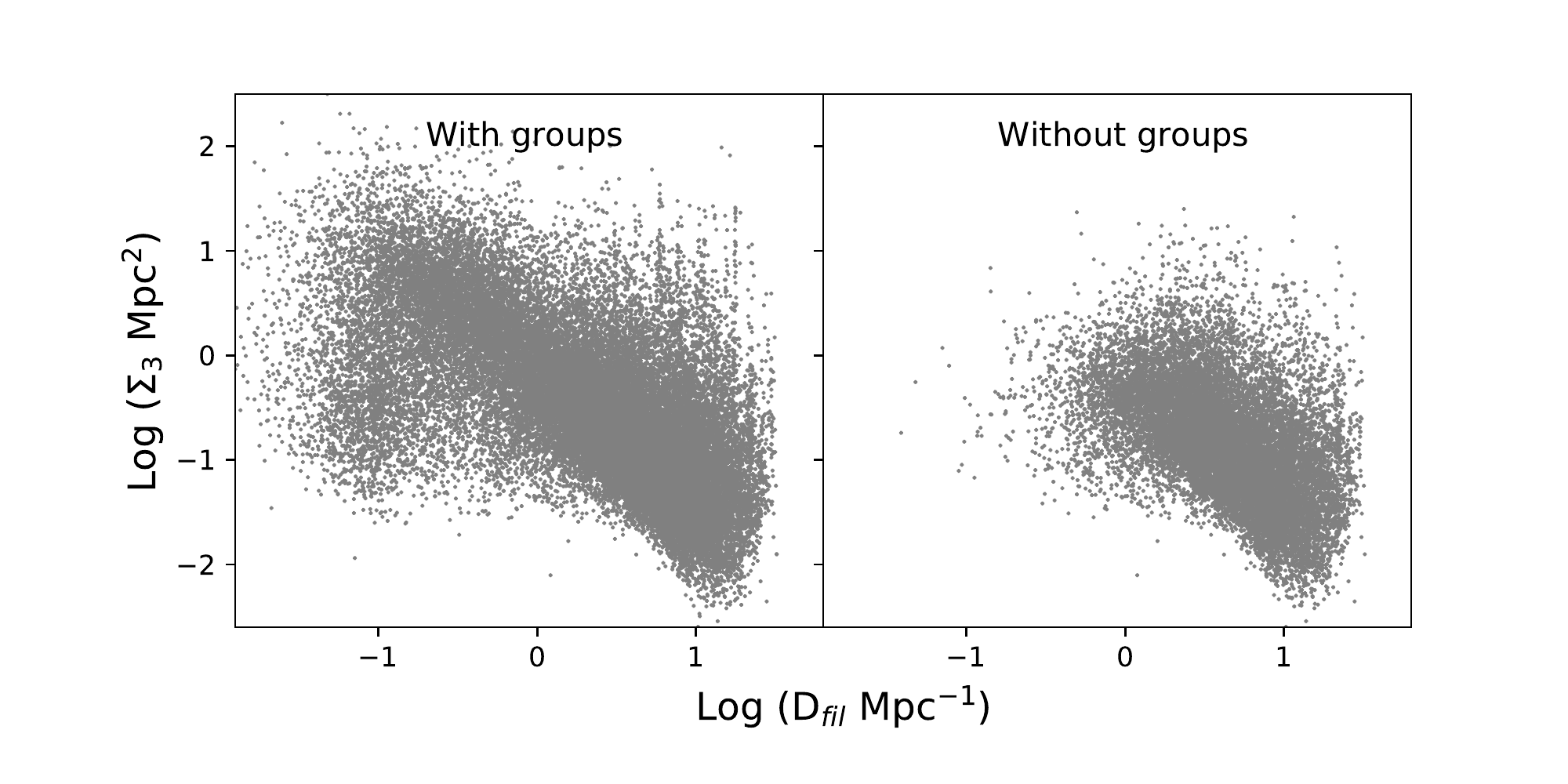}
	\setlength{\abovecaptionskip}{-10pt} 
	\caption{Relationship between local density $\Sigma_3$ and distance from the filament spine D$_{fil}$ for all the galaxies (left) and for only galaxies that are not in groups (right). An exception to the approximately inverse relationship between Log $\Sigma_3$ and Log D$_{fil}$ is the concentration of galaxies at moderate density but D$_{fil} \sim -1$;  these galaxies are in groups that are relatively small but that define the filament spine.}
	\vspace{10mm}
\end{figure*}

The redshift distortion along the line of sight caused by orbital motions presents a particular limitation in defining the distance to filaments for galaxies that are in clusters and large groups.  Our estimate of D$_{fil}$ uses distances based on the redshift of the group.  In many cases, these group centers were part of the filament spine, so D$_{fil}$ is systematically underestimated.  Assuming that the component of the galaxy displacement from the group center along our line of sight is, on average, the same as each component of the displacement on the sky, our estimate of D$_{fil}$ for these group galaxies is, on average, $\sqrt{2/3}$ or about $80\%$ of its actual value.  

Because groups provide a dynamically different environment that might obscure the environmental effect of filaments, most previous observational studies of filaments exclude group galaxies from their samples. In this paper we present results for both the full sample and a sample that excludes all group galaxies.  The right panel of Figure 5 shows the 
relationship between Log D$_{fil}$ and Log $\Sigma_3$ for this sample, which spans a local density roughly up to Log $\Sigma_3 \sim 0.0$, or $\Sigma_3 \sim 1$ Mpc$^{-2}$ and  distance to the filament spine down to roughly LogD$_{fil} \sim 0.0$, or D$_{fil} \sim 1$ Mpc. 

\section{Results}

Our goal is to compare galaxies that are very similar except for their large-scale environment. 
We consider the environmental dependence of three properties: stellar mass, color at fixed stellar mass, and HI deficiency at fixed stellar mass.  In the first subsection below, we show the dependence of these properties on large-scale structure and on local density independently.  In the following subsection we address their dependence on large-scale structure and local density simultaneously, to see if there is a dependence on large-scale structure once local density is taken into account.  We then consider the dependence of HI deficiency on environment when both stellar mass and color are fixed.

Both color and HI deficiency vary systematically with stellar mass. One practical approach to comparing galaxies at fixed stellar mass is to split the sample into mass bins that are small enough to minimize differences in mass but large enough to provide useful statistics.  Another is to fit the dependence on mass to a function and consider the residuals around that function.  In this paper we adopt the latter approach.  Besides using residuals to ``correct" for  mass in this way, we use them to correct for local density and color in the subsections that follow.  We also include heliocentric distance as an independent parameter, since our main sample is not complete over the sample volume; this takes into account trends that might result from environmental conditions that vary with distance, and therefore with flux limit.

Figure 6 illustrates our process of using residuals from multiple regressions.  In this case, HI deficiency is fit as a function of three independent variables: heliocentric distance, log stellar mass Log(M$_\star$ M$_{\odot}$$^{-1}$), and perpendicular distance from the filament spine D$_{fil}$.  The purpose is to find functions that remove the dependence on distance and stellar mass, leaving only the residual dependence on the environmental variable of interest, the distance to the filament spine.

\begin{figure*}
	\epsscale{1.0}
	\includegraphics*[width=500pt]{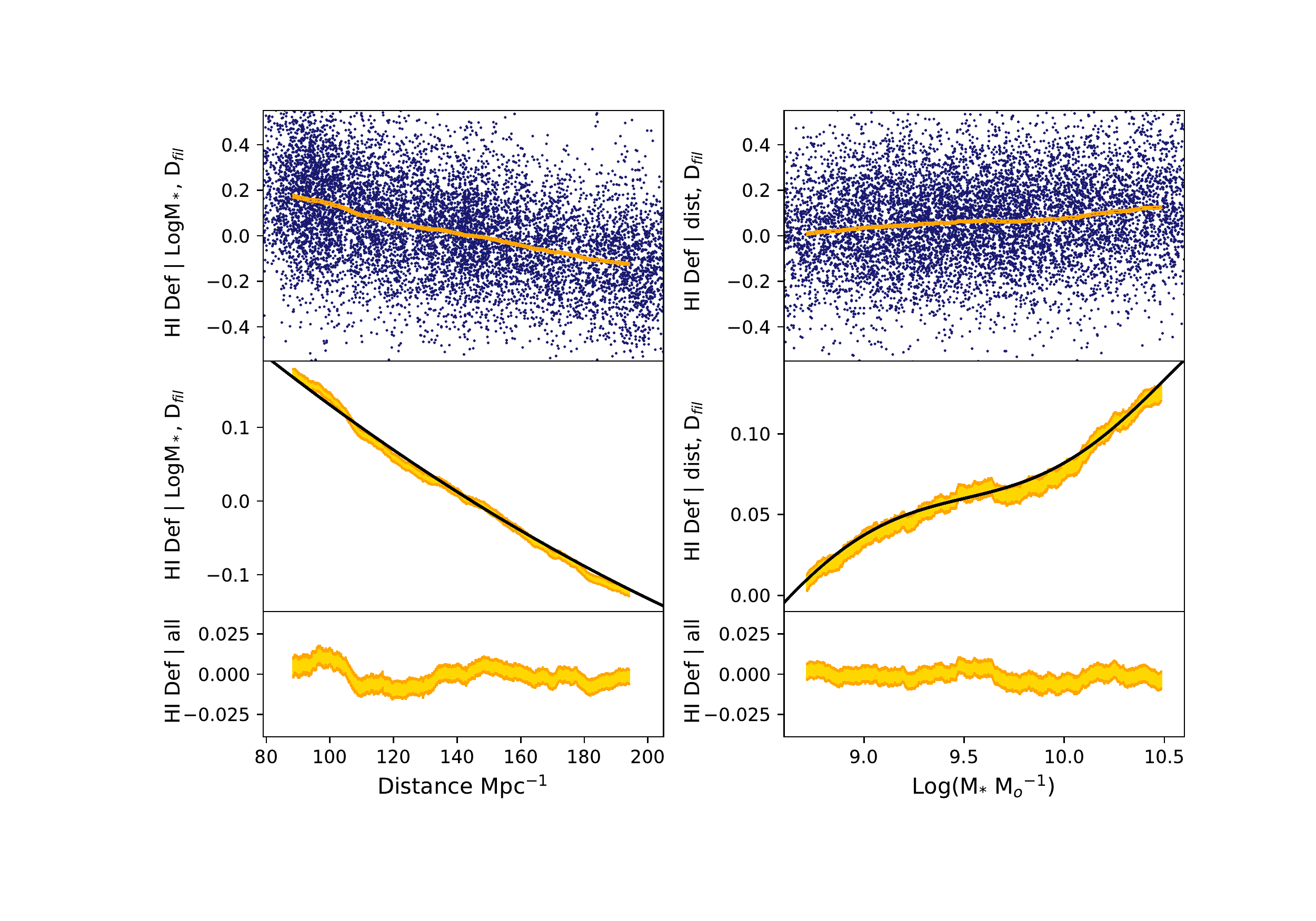}
	\setlength{\abovecaptionskip}{-25pt} 
	\setlength{\belowcaptionskip}{0pt} 
	\caption{Residuals for HI deficiency, illustrating the process for estimating HI deficiency as a function of the perpendicular distance D$_{fil}$ to the nearest filament spine, at fixed heliocentric distance and stellar mass.  The top two rows show partial residuals based on regressions for HI deficiency as a function of distance, log stellar mass Log M$_\star$, and D$_{fil}$. The left column shows residuals relative to only Log M$_\star$ and D$_{fil}$, leaving the remaining dependence on distance.  The right column shows residuals relative to only distance and D$_{fil}$, leaving the remaining dependence on mass.  In the top row, blue dots indicate individual galaxies, while orange lines indicate the mean value with N=2000 boxcar smoothing. The width of the orange lines gives the standard error in the mean. The second row is a closeup of the same smoothed residuals, with black lines showing the best polynomial fits to the unsmoothed data. The bottom row shows the full residuals, where the dependences on both of these variables have been successfully removed. The quantity of interest, the HI deficiency as a function of D$_{fil}$ at fixed distance and stellar mass, is the residual relative to distance and Log M$_\star$ leaving the remaining dependence on D$_{fil}$. This is shown in Figure 7.}
\end{figure*}

The top row shows the approximately linear dependence of HI deficiency on distance and on 
Log(M$_\star$ M$_{\odot}$$^{-1}$), although with considerable scatter. Individual galaxies are shown as blue dots, and the mean value with boxcar smoothing over N=2000 is superimposed in orange.  Each panel shows the dependence on just one independent variable; the left panel, for example, shows the partial residuals of HI deficiency relative to Log M$_\star$
and D$_{fil}$, with only the dependence on heliocentric distance remaining. 

Despite their approximately linear relationships, some of the data warrant slightly more complicated functions.  Our procedure is to fit to a polynomial with up to ten terms and choose the best one according to the Bayesian information criterion (BIC) criterion (Schwarz 1978).  
The BIC is based on the likelihood function as well as the sample size and the number of free parameters.  It is conservative in that it imposes a higher penalty for additional terms than other commonly used statistical tests for determining best models, including the adjusted $R^{2}$ and the Akaike information criterion (AIC).  In nearly all cases, the best function is quite simple, with 1-3 terms.  The second row of Figure 6 is a closeup of the top row showing the smoothed values along with the fitted function. The bottom row shows the full residual, with the dependences on all three independent variables removed.  The quantity of interest, the HI deficiency as a function of D$_{fil}$ at fixed distance and stellar mass, is the residual relative to distance and LogM$_\star$ leaving the remaining dependence on D$_{fil}$; this is shown in the bottom left panel of Figure 7.  (Note that the uncertainties indicated by the width of the orange lines are correlated across neighboring points because of the smoothing.)  

The example in Figure 6 specifically illustrates the process for finding the dependence of HI deficiency on the distance to the filament spine using residuals relative to heliocentric distance and stellar mass.  The same process is applied in the results below to determine how other galaxy properties depend on other environment variables. 
We emphasize that this method is a practical heuristic to see if there are residual dependences on environment. It is not a method that provides intrinsic physical relationships among the variables, since it depends on our selection criteria.  Other techniques (e.g. those in Toribio et al. 2011) would be used to find intrinsic physical relationships. 

An important related point is that the residuals for any quantity are scattered around zero. The partial residuals for color and HI deficiency, for example, do not correspond to the actual measured values for color and HI deficiency, but to the \textit{difference} between the measured values and the best fit at that heliocentric distance and stellar mass.

In the results that follow, values for HI deficiency residuals are for galaxies detected in HI unless otherwise stated; some statistical comparisons include detection limits as well, and are noted explicitly.  

\subsection{Dependence on large-scale structure and local density independently}

In this subsection, we consider the environmental dependence of galaxy properties on large-scale structure and local density independently. 

Figure 7 shows smoothed partial residuals for stellar mass at fixed distance, g-i color at fixed distance and stellar mass, and HI deficiency at fixed distance and stellar mass, as functions of distance from the filament spine D$_{fil}$ and as functions of Log D$_{fil}$.  Results include  both the full sample (light orange) and the smaller sample with groups removed (dark blue).  All three properties depend on distance to the filament: in every case, the residuals show a slope that is statistically distinguishable from zero. 

\begin{figure*}
	\epsscale{1.0}
	\includegraphics*[width=400pt]{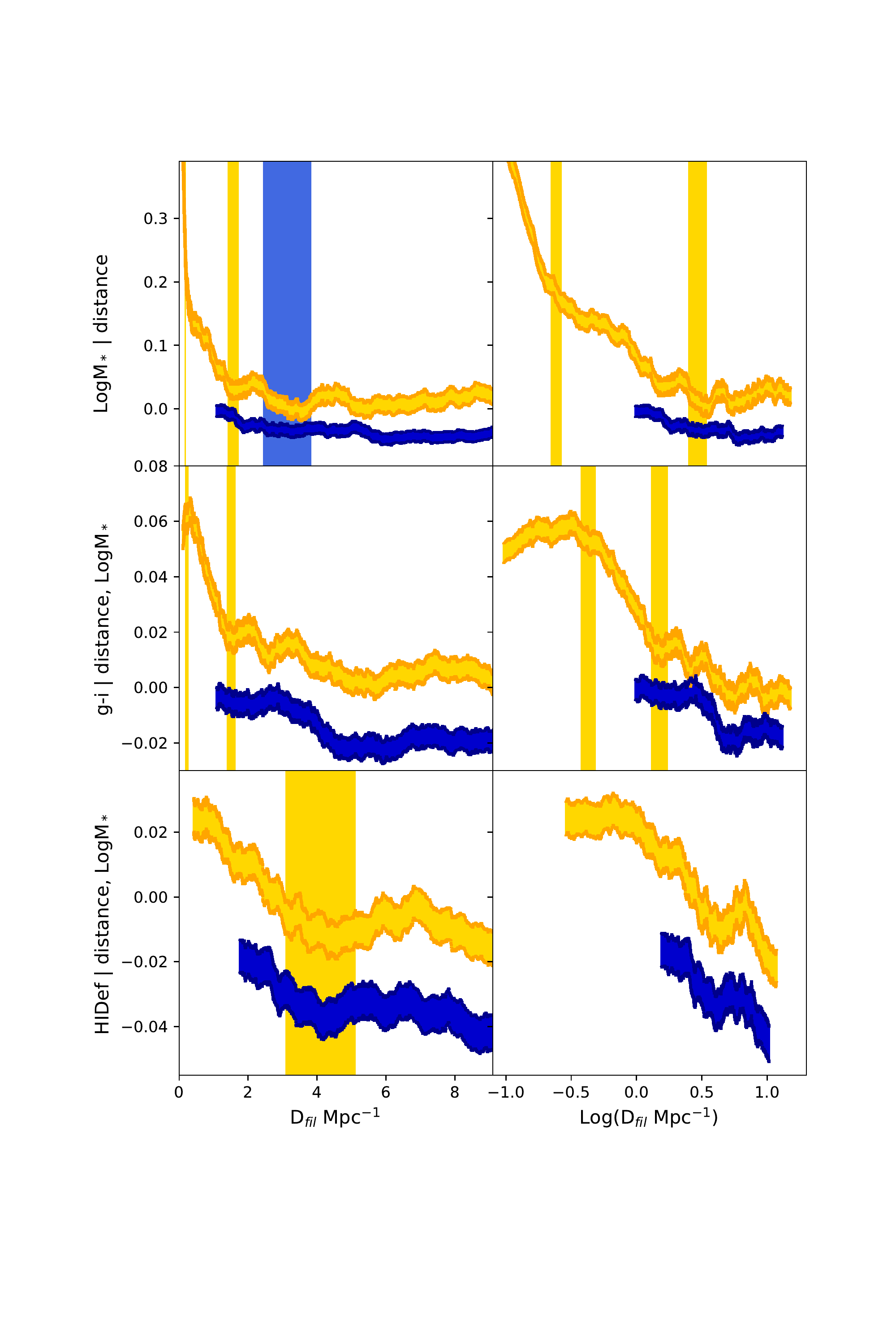}
	\setlength{\abovecaptionskip}{-70pt} 
	\caption{Smoothed partial residuals for stellar mass at fixed heliocentric distance, g-i color at fixed stellar mass and distance, and HI deficiency at fixed stellar mass and distance, as functions of distance from the filament spine D$_{fil}$ and Log D$_{fil}$.  Orange lines show residuals including all galaxies while dark blue lines show residuals when galaxies in groups are excluded. Smoothing is over N=2000 galaxies, and the thickness of the lines gives the standard error in the mean. The blue lines are shifted slightly downward for clarity, so they do not overlap with the orange lines. Vertical lines give the locations of all statistically significant changes in slope based on the unsmoothed data.  Only one change in slope (the wide blue vertical line in the first panel) is found for the sample without groups; the lighter vertical yellow lines apply to the sample that includes groups.  All three galaxy properties decrease with distance from the filament spine.  If considered as function of D$_{fil}$, all three properties show a transition to a flatter slope at 1.5-4.0 Mpc. If considered as a function of Log D$_{fil}$, both stellar mass and g-i still show a transition to a flatter slope at a similar value for D$_{fil}$, but HI deficiency does not.}
\end{figure*}

A particular question is whether filaments create a threshold in the availability of cold gas, for example because the galaxy is ``dynamically detached" from the cosmic web (Aragon-Calvo 2016). Such a threshold could appear as a change in slope in galaxy properties as a function of D$_{fil}$.  Figure 7 shows all transitions that are significant at the 95\% level in the unsmoothed data according a Davies test for a change in slope (Davies 1987), with locations and standard errors determined with the segmentation package in the R statistics environment.  Transitions for the full sample are given by light yellow vertical bands, while transitions for the sample without groups are given by darker blue vertical bands. (There is only one of the latter, the wide band in the upper left panel.)  

In interpreting these figures, it is important to keep in mind that values are correlated across neighboring points through boxcar smoothing. While the smoothing process allows us to visually examine features that would otherwise be hidden in noise, this artificial correlation makes it difficult to intuitively judge statistical significance.  All statistical tests for slopes and linearity are performed on the \textit{unsmoothed} data.

Considering the left column of Figure 7, we see that for the full sample of galaxies, all three properties do indeed show a transition to a flatter slope at 1.5-4.0 Mpc.  The transitions for Log M$\star$, and g-i color (narrow yellow vertical bands) occur at about 1.5 Mpc, while the transition for HI deficiency occurs slightly farther out, at about 4 Mpc (although with a large uncertainty).  This could be interpreted as an abrupt decrease in the supply of gas that produces a delayed transition in reddening as galaxies are pulled closer to the filament spine.  Note, however, that for the smaller sample without groups there is no statistically significant transition in slope. 

At very small values of D$_{fil}$, about 0.2 Mpc, both LogM$\star$, and g-i color show additional transitions:  the slope for mass is much steeper at small D$_{fil}$, while the slope for color is positive instead of negative.  Galaxies at very small values of D$_{fil}$ are typically members of groups that are nodes in the minimal spanning tree that define the filament.  These are not necessarily very large groups, but they must have at least one large central galaxy to be identified in the group catalog. In this sense the behavior at very small D$_{fil}$ is dependent on the specific way the filament spine was determined.

The appearance of transitions in slope are affected by the fact that D$_{fil}$ for groups that are nodes in the filament will be underestimated (on average, about 80\% of their true value, as described above). 
Relocating these galaxies to slightly higher D$_{fil}$ could make the changes in slope less dramatic and could make them appear at slightly higher D$_{fil}$.  However, shifts this small would not be enough to fully explain the existence of the transitions, considering that the flat part of the curves extend further than 10 Mpc.

If we consider galaxy properties as a function of the logarithm of D$_{fil}$ (right column), these changes in slope are less dramatic and, for HI deficiency, not significant at the 95\% level.  The transition from a steep negative slope in Log M$_{\star}$ at Log D$_{fil} \sim -0.6$ might be explained by the $\sim 80\%$ shift in D$_{fil}$ for groups galaxies, which would push some galaxies at low D$_{fil}$ out to the inflection point at Log D$_{fil}\sim -0.5$. 
One interpretation of the simpler, nearly linear relationships is that Log D$_{fil}$ is a more natural parameter to use than D$_{fil}$, perhaps because of its roughly inverse relationship to Log $\Sigma_3$.   

Even in the cases with no \textit{transition} in slope, all the curves in Figure 7 are statistically distinguishable from zero, indicating a dependence on distance from the filament spine.

An alternative way to assess the dependence of galaxy properties on large-scale structure is through the categories of group, filament, cluster, and void.  Figure 8 compares the same three properties --- stellar mass, color at fixed stellar mass, and HI deficiency at fixed stellar mass --- in these different large-scale environment categories.  The average values for void galaxies are used as a reference point (the dashed blue line).  The points with error bars show the average values for groups, filaments, and clusters relative to this reference point.  

In general, the average values for each property decrease from group to tendril environments. In particular, the values for filament galaxies are higher than either tendril or void galaxies, indicating larger, redder, and more HI deficient galaxies. However, tendril galaxies have a similar color and HI deficiency to void galaxies, despite being in denser environments and having more massive galaxies.

\begin{figure}
	\epsscale{1.0}
	\vspace{-0mm}
	\includegraphics*[width=250pt]{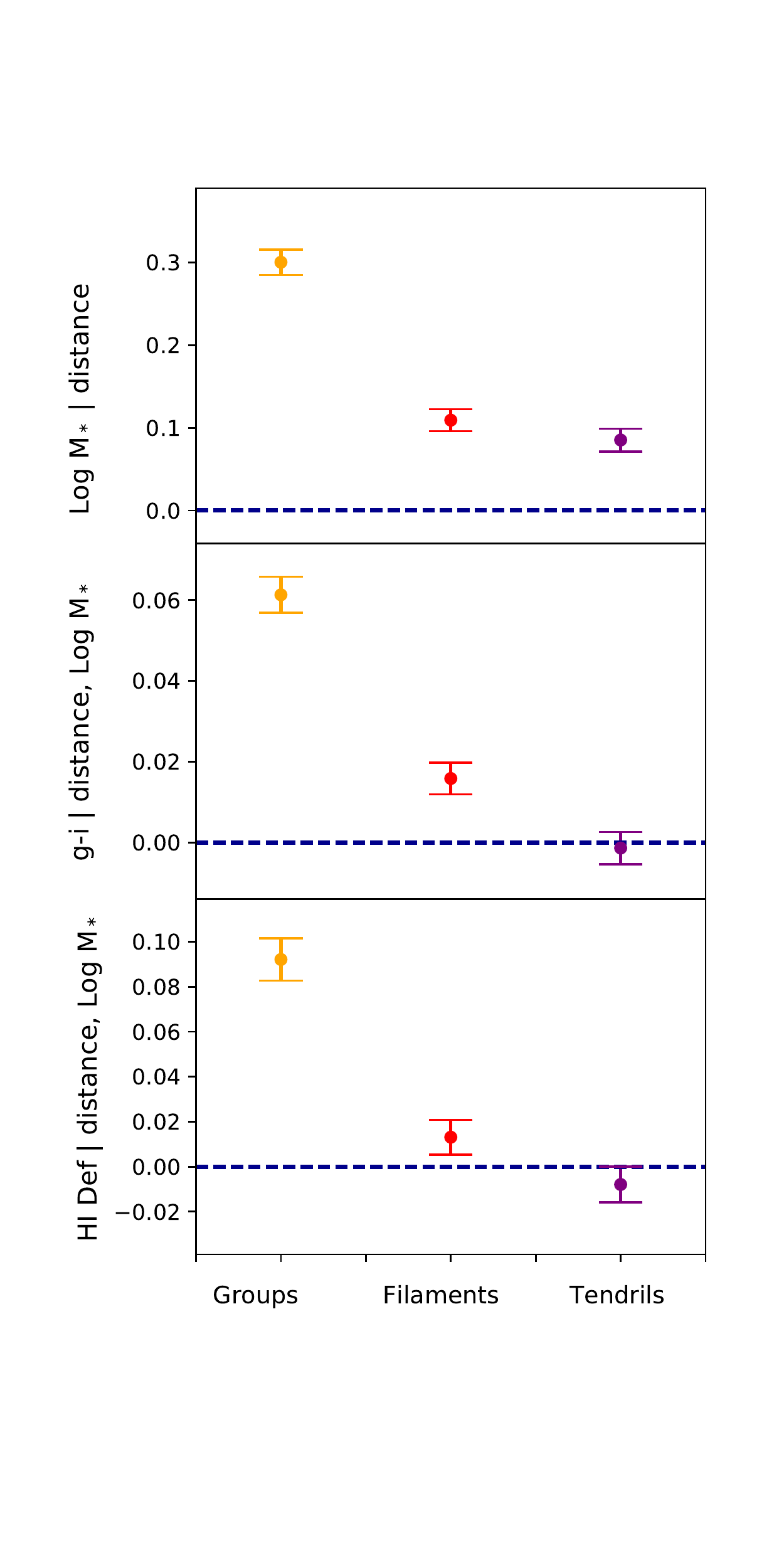}
	\setlength{\abovecaptionskip}{-70pt} 
	\caption{Comparison of stellar mass, color at fixed stellar mass, and HI deficiency at fixed stellar mass in different large-scale environments.  The averages for galaxies in groups, filaments, and tendrils (left to right) is compared with the average for galaxies in voids (horizontal dashed line).  Error bars give the standard error for the difference between each category mean and the mean for void galaxies. In general, these values decrease from group to tendril environments, as one might expect based on local density.  However, tendril galaxies have a similar color and HI deficiency as void galaxies, despite being in denser environments.}  
\end{figure}

Finally, we consider the dependence of these properties on the local environment as opposed to the large-scale environment.  Figure 9 shows the same properties as in Figure 7, but as a function of Log $\Sigma_3$.  All three galaxy properties increase with local density.  There is only one transition in slope: the slope for g-i color is flat for both samples at low local density.  The fact that these relationships are simpler than those in Figure 7 suggests that local density is the primary environmental driver. In the next subsection we consider whether we should consider local density as the \textit{only} environmental driver, or whether large-scale structure has an impact at fixed local density.

\begin{figure}
	\epsscale{1.0}
	\vspace{-0mm}
	\includegraphics*[width=250pt]{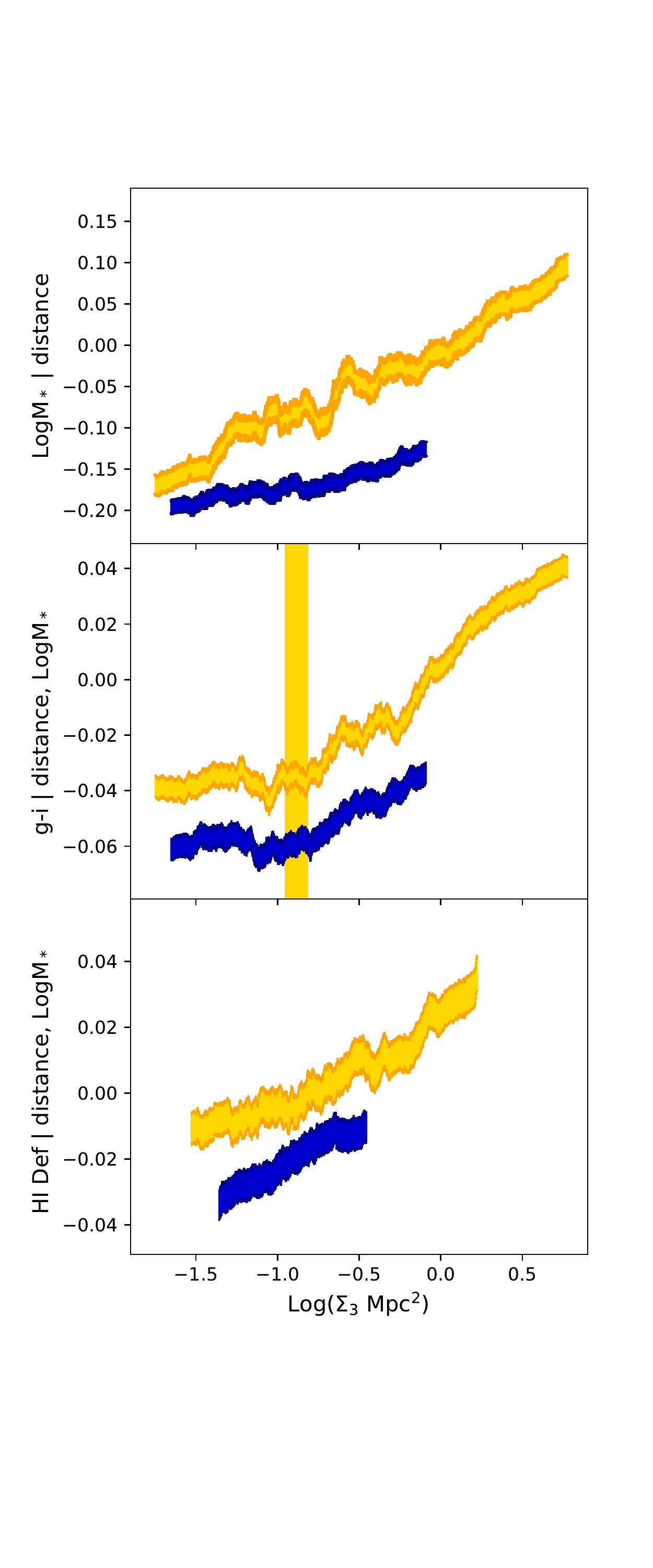}
	\setlength{\abovecaptionskip}{-130pt} 
	\caption{Like Figure 7, but as a function of log local density.  All three galaxy properties increase with local density.  There is only one transition in slope: the slope for g-i color is flat for both samples at low local density.}
	\vspace{-15mm}
\end{figure}

\subsection{Does large-scale structure have an impact at fixed local density?}

The previous section shows that mass, color at fixed mass, and HI deficiency at fixed mass depend on large-scale environment, and also on local density.  Here we consider the question of whether models that include both local and large-scale measures of environment simultaneously
provide improved fits over models that include either one individually.

First we compare fits that include LogD$_{fil}$ alone, Log$\Sigma_3$ alone, or both together. 
As we did above in the context of polynomial fits, we use the BIC statistic to determine which of these models are better.  When the full galaxy sample is included, models that include both are superior for all three galaxy properties: stellar mass, color at fixed stellar mass, and HI deficiency.  Recall that the BIC imposes a strong penalty for additional free parameters; other common tests, including the AIC and $R^{2}$$_{adj}$, prefer the models with both Log D$_{fil}$ and Log $\Sigma_3$ even more than the BIC does.

When group galaxies are excluded from the sample, models for stellar mass and color that include both Log D$_{fil}$ and Log $\Sigma_3$ are superior according to the AIC and $R^{2}$$_{adj}$ criteria, while the BIC favors local density alone. Models for HI deficiency incorporating non-detections that include both Log D$_{fil}$ and Log $\Sigma_3$ are superior according to all three criteria, while models for HI deficiency based on detections alone that include both Log D$_{fil}$ and Log $\Sigma_3$ are superior according to the AIC and $R^{2}$$_{adj}$ criteria, while the BIC favors local density alone.   

Considering the models where large-scale structure is described in terms of the group, filament, tendril and void categories, we find that including both local density and large-scale structure provides improved fits over either local density or large-scale category alone, according to all three criteria. 

In this paper, we are particularly interested in the extent to which fits for HI deficiency are improved when large-scale structure information is included in addition to local density.  The effect we see here is strong.  For fits to Log D$_{fil}$ the models that include both are improved by $\Delta BIC=17$ (including group galaxies) and $\Delta BIC=6$ (excluding group galaxies). For fits to large-scale structure categories, the models that include both large-scale structure and local density are improved by $\Delta BIC=43$.  These values are generally considered ``strong" to ``very strong" evidence that the model with the lowest BIC is indeed statistically favorable (e.g. Raftery 1995).

Figure 10 shows the remaining dependence of galaxy properties on Log D$_{fil}$ when Log $\Sigma_3$ is included as an additional independent variable.  For the full sample (light orange), there is still a negative slope out to Log D$_{fil} \sim 0.5$. (The transition in slope at small D$_{fil}$ could again be caused by the $80\%$ shift in D$_{fil}$ for group galaxies). 
At larger distances of D$_{fil} > 0.5$, or Log D$_{fil} >$ 3 Mpc, the slope is positive (with a probability p=0.004 that it is statistically consistent with zero).  For the sample without groups (dark blue), the transition at Log D$_{fil} \sim 0.5$ appears only at the p=0.11 confidence level, and the slope at higher D$_{fil}$ is positive only at the p=0.07 level. 

\begin{figure}
	\epsscale{1.0}
	\includegraphics*[width=250pt]{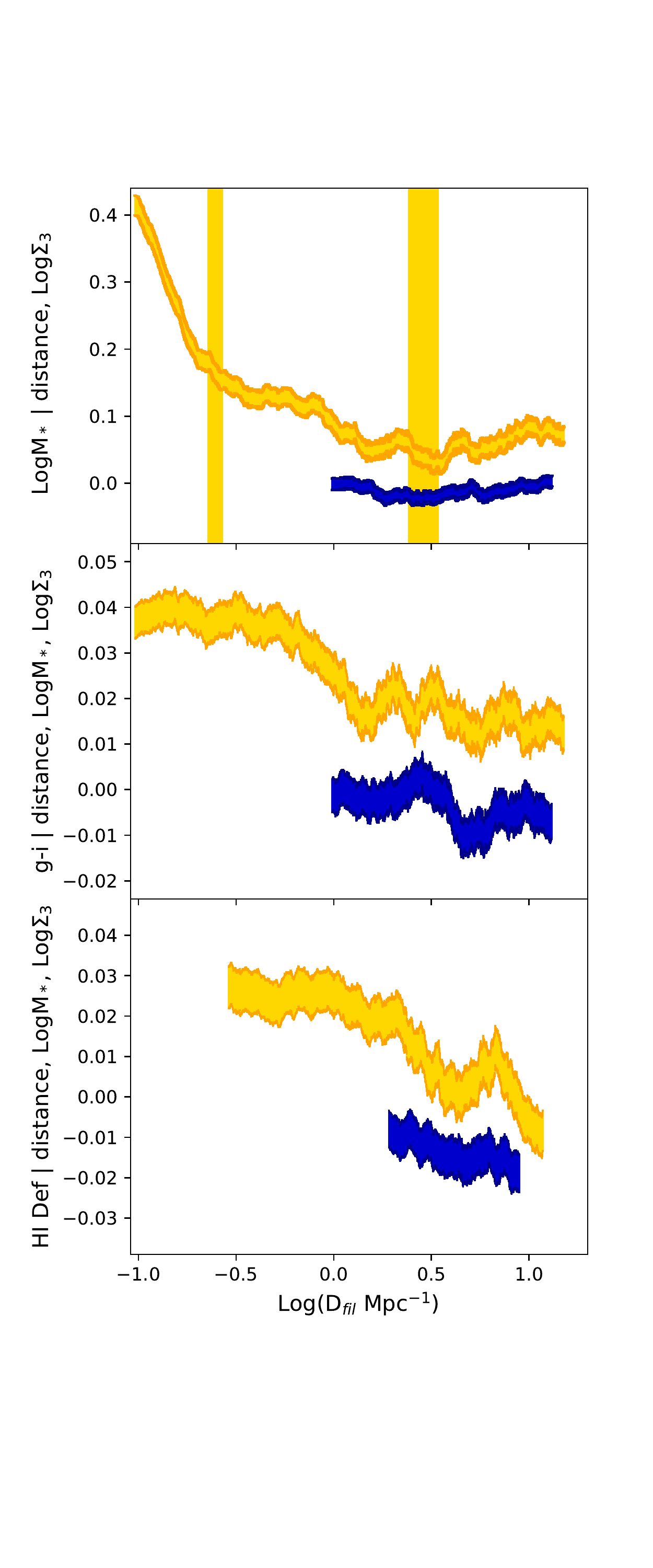}
	\setlength{\abovecaptionskip}{-45mm} 
	\caption{Like the second column of Figure 7, but for fits that include local density as an additional independent variable.  At fixed local density, there is still a statistically significant overall decrease with LogD$_{fil}$ at small values of LogD$_{fil}$ for the full sample (orange) and, in the case of HI deficiency, for both samples.}
	\vspace{-12mm}
\end{figure}

The curves for color and HI deficiency in Figure 10 are all statistically consistent with straight lines, and all have negative slopes that are significantly different from zero, with the exception of color in the sample without groups, which is different from zero at the p=0.14 level only.  Of particular interest to us is the negative slope for HI deficiency even in the sample without groups, which is significant at the p=0.014 level if only HI detections are used and at the p=0.0076 level if the regressions use survival analysis techniques to include HI flux limits for non-detections.  

Figure 11 shows the dependence on large-scale structure at fixed local density in terms of the group, filament, tendril, and void categories.  As in Figure 9, the average values of void galaxies are used as a reference (the dashed blue line), while the points with error bars show  the differences in average values for groups, filaments, and clusters relative to this reference value.  

\begin{figure}
	\epsscale{1.0}
	\includegraphics*[width=250pt]{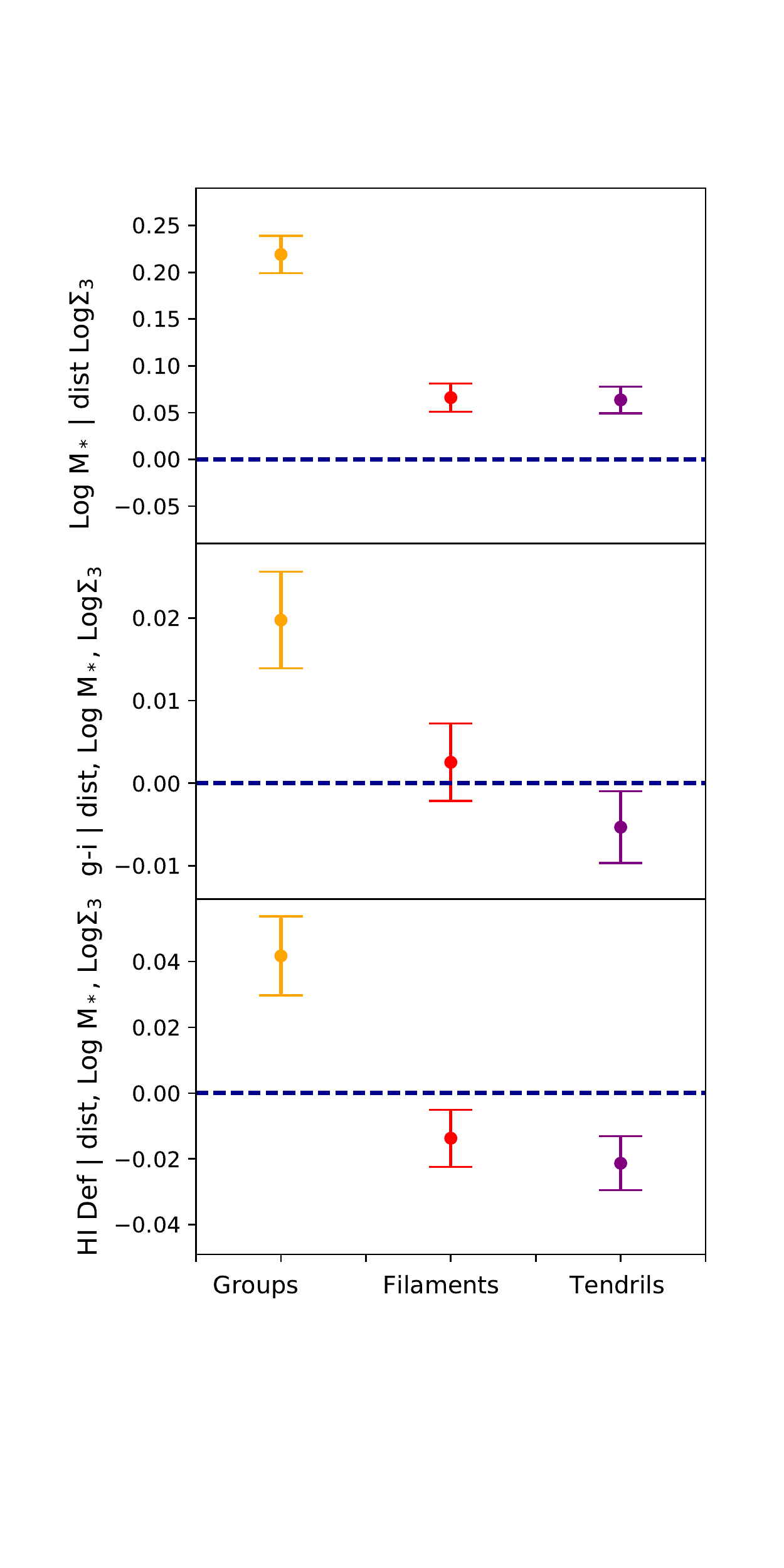}
	\setlength{\abovecaptionskip}{-35mm} 
	\caption{Like Figure 9, but for fits that include local density as an additional independent variable.  Even at fixed local density, there is still a general decrease in each of these properties from groups to tendrils (left to right).  However, for their density, galaxies in tendrils are bluer and more HI deficient than galaxies in voids (and are therefore below the horizontal dashed line.)} 
	\vspace{-5mm}
\end{figure}

Separating the galaxy populations by large-scale structure category shows a complementary picture to the one in Figure 10.  Void galaxies are both far from filament spines and (unlike some of the galaxies at high D$_{fil}$) are not part of structures like groups or tendrils. With these structures removed, we see that void galaxies still have smaller masses than any other category, even at fixed density. Considering color and HI deficiency at fixed density we see that only group galaxies are significantly more red than void galaxies, and that tendril galaxies are  \textit{less} HI deficient than void galaxies for their density (p=0.07 for HI detections only, p=0.009 including HI flux limits) 

In summary, including information about large-scale structure along with local density provides superior fits to either one alone.  When all galaxies are included in the sample, all three properties still decrease overall with Log D$_{fil}$, and even when groups are excluded HI deficiency still decreases with Log D$_{fil}$.  Similarly, there is still an overall trend for the properties to decrease across the categories group, filament, and tendril.  
However, tendril galaxies are slightly less deficient than void galaxies for their density despite having higher stellar masses for their density.  Tendrils may be regions where the processes that remove HI are not efficient, or where processes that replenish HI can compensate for them.

\subsection{Is there evidence that a loss of HI gas precedes reddening?}

Finally we consider the connection between color and HI deficiency.  If galaxies entering filaments (or, more generally, entering increasingly dense environments) first lose HI gas and then redden afterward because they stop forming new stars, we might see that galaxies with the same mass \textit{and color} are more HI deficient closer to the filament spine.  In other words, galaxies that are otherwise the same may have their HI gas supply rapidly removed because they have entered the new environment, while their other properties have not yet changed.

To look for this phenomenon, we include g-i color as an independent variable along with stellar mass in fits for HI deficiency as a function of environment. We find that, indeed, including color along with either Log D$_{fil}$  or Log $\Sigma_3$ improves the fit according to the BIC, AIC, and reduced $R^{2}$$_{adj}$ criteria.  (Note that HI deficiency is strongly correlated with red color.) Similarly, including color along with the categories of group, filament, tendril, or void improves the fits according to all three criteria. 

Figure 12 shows that at fixed color and stellar mass, HI deficiency is still higher at low D$_{fil}$ and high $\Sigma_3$. Figure 13 shows that at fixed color and stellar mass, HI deficiency still tends to decrease from groups to tendrils. This fits a picture where, as galaxies enter denser regions, they first lose HI gas and then redden as star formation is reduced. 

\begin{figure*}
	\epsscale{1.0}
	\includegraphics*[width=500pt]{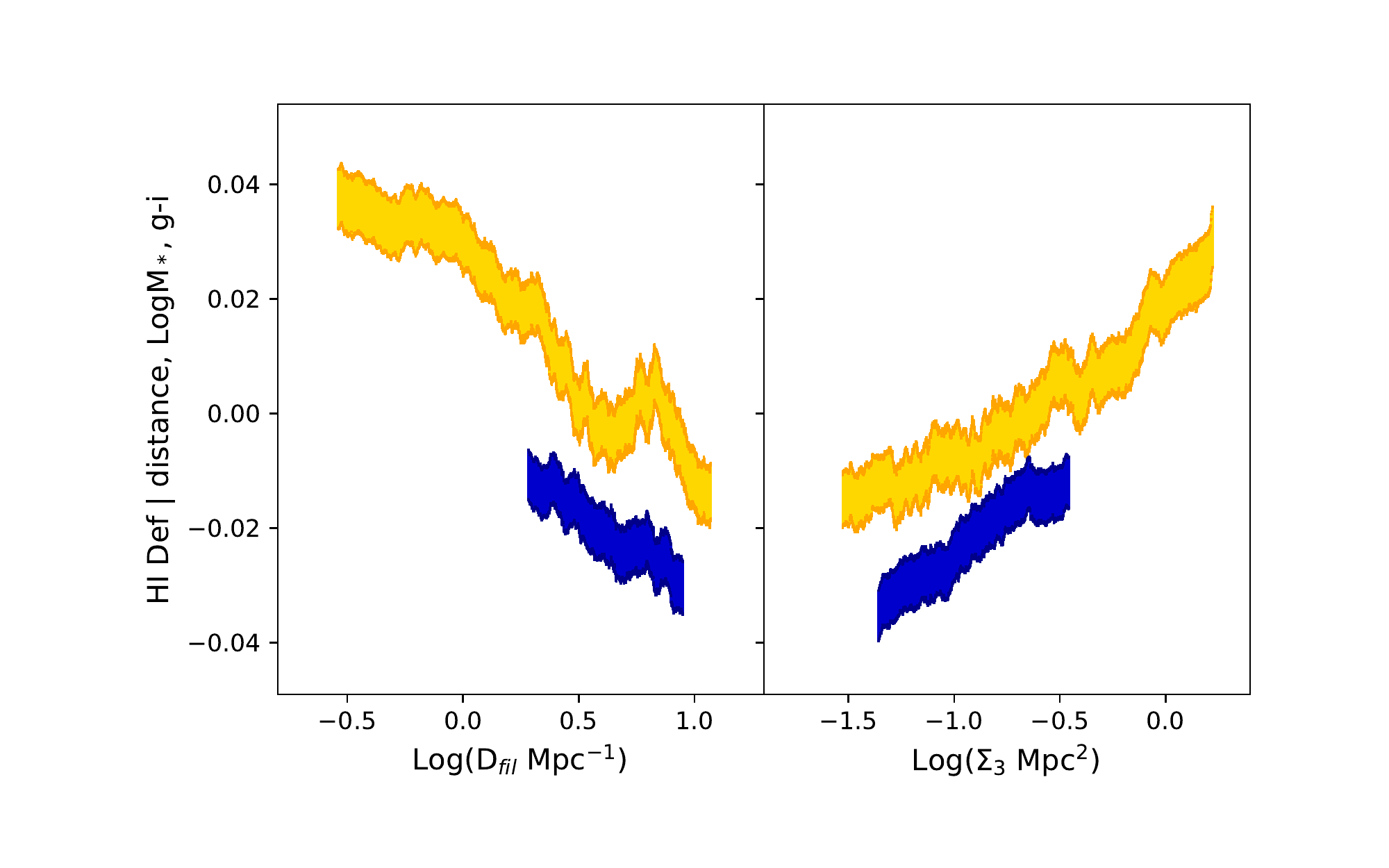}	\setlength{\abovecaptionskip}{-3mm} 
	\caption{Like Figure 7, but for fits to HI deficiency as a function of LogD$_{fil}$ and Log$\Sigma_3$ that include g-i color as an additional independent variable.  At fixed color and stellar mass, HI deficiency is still higher at low D$_{fil}$ and high $\Sigma_3$. These trends support a scenario where the removal of HI gas upon entering a denser region precedes reddening.}
	\vspace*{10mm}
\end{figure*}

\begin{figure}
	\epsscale{1.0}
	\includegraphics*[width=250pt]{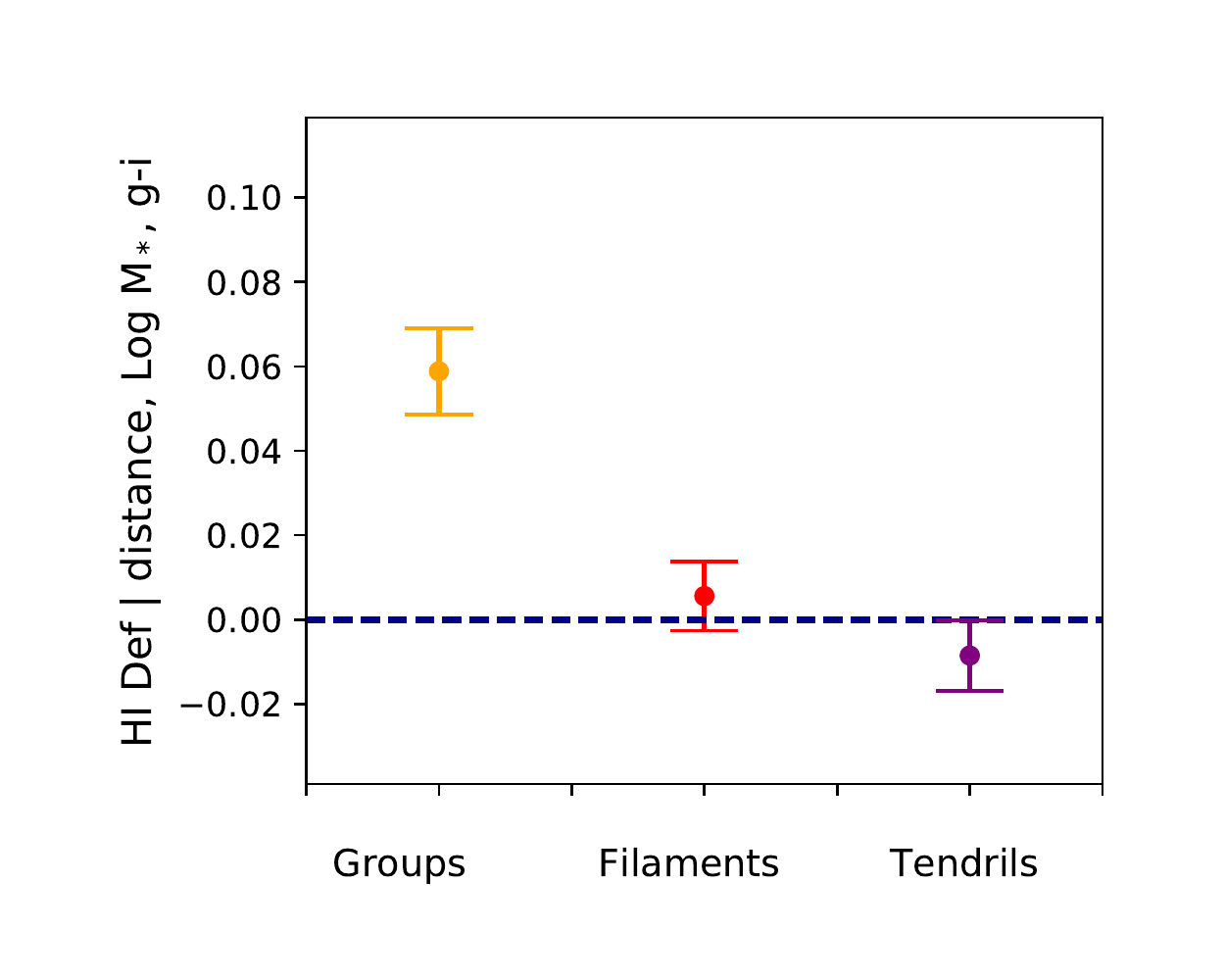}
	\setlength{\abovecaptionskip}{-7mm} 
	\caption{Like Figure 9, but for fits to HI deficiency that include g-i color as an additional independent variable. At fixed color and stellar mass, HI deficiency still tends to decrease from groups to tendrils (left to right).}
	\vspace{-0mm}
\end{figure}

\vspace{3mm}
\section{Conclusions and Discussion}

Our primary results are the following: 

1) The HI deficiency of galaxies at fixed stellar mass, along with color at fixed stellar mass and with stellar mass itself, decreases within a few megaparsecs of the filament spine D$_{fil}$ and then flattens out.  This change in the slope of HI deficiency could be interpreted as an abrupt reduction in available atomic hydrogen gas as galaxies enter the region of a filament, for example as hydrogen gas is heated or as galaxies are dynamically disconnected from the cosmic web of cool gas.  
While the slope for HI deficiency at fixed stellar mass as a function of the distance to the filament spine changes at about 4 Mpc, the slope for galaxy color at fixed stellar mass changes at a smaller distance of about 1.5 Mpc.  This could be interpreted as reddening after gas loss as the galaxies move closer to the filament spine.  If galaxy properties are considered as a function of Log D$_{fil}$ instead of D$_{fil}$, these changes in slope are less dramatic, and are not statistically significant in the case of HI deficiency. However, there is still a nonzero slope indicating a smooth dependence on environment. One interpretation of this simpler, nearly linear relationships is that Log D$_{fil}$ is a more natural parameter to use than D$_{fil}$, perhaps because of its roughly inverse relationship to Log $\Sigma_3$. 

2) HI deficiency at fixed stellar mass, along with color at fixed stellar mass and stellar mass itself, depends on both large-scale structure and local density.  Regressions that include both local density and distance from the filament spine give improved fits over either environment variable alone, according to BIC, AIC, and adjusted $R^{2}$ criteria.  When group galaxies are excluded from the sample, regressions that include both are improved according to the AIC and $R^{2}$ criteria; the BIC criterion, which imposes a stronger penalty for including more variables, favors local density alone for stellar mass and color, but again favors including both large-scale structure and local density for HI deficiency.  Regressions that include both local density and the large-scale structure categories of group, filaments, tendrils and void provide improved fits over either local density or large-scale category alone, according to all three criteria. 

3) When group galaxies are excluded, and when local density is taken into account, we do not detect a dependence of stellar mass on distance from filament spines.  However, when separated into the categories of group, filament, tendril, and void, we detect higher masses at fixed local density for filament and tendril galaxies relative to void galaxies. (The average for group galaxies, meanwhile, is considerably higher than for any other category.)

4) Although galaxies in tendrils are in significantly denser environments, on average, than galaxies in voids, they are not redder or more HI deficient. Indeed, for their density, they are bluer and less HI deficient than void galaxies.  This stands in contrast to the fact that galaxies in tendrils are more massive than those in voids, suggesting a more advanced stage of evolution. Tendrils may be regions where the processes that remove HI are not efficient, or where processes that replenish HI can compensate for them.  

5) The galaxies in our sample show a decrease in HI deficiency with distance from the filament spine, even at constant local density.  This suggests that the cold gas supply is reduced near filaments.  However, the HI deficiency for galaxies in the filament category is, at constant local density, not distinguishable from that in voids if tendrils are removed from the void as a separate category, since it is the tendril category that is the least HI deficient for constant density.  In comparing these results to those of Kleiner et al. (2017), who find evidence of replenishment of cold gas near filaments, we note that most of our galaxies are in the stellar mass range $8.5<$log(M$_{*}$/M$_\odot)<10.5$ (see the right column of Figure 6), whereas they see replenishment only for large galaxies with log(M$_{*}$/M$_\odot)>11$.  We also note, however, that our average deficiency for filaments is indeed slightly (although less than 2$\sigma$) below that of voids once tendrils are removed. 

6) At fixed stellar mass \textit{and color}, galaxies closer to the filament spine (or in high density environments) are more deficient in HI.  This fits a picture where, as galaxies enter denser regions, they first lose HI gas and then redden as star formation is reduced. 

We note that the differences we present here in HI deficiency and color across environments are relatively small: about 0.10 and less for HI deficiency and 0.06 and less for color.  Larger differences have been seen previously when comparing high-density regions to moderate-density regions  -- for example, when comparing clusters to their surrounding regions (e.g. Solanes et al. 2001). Because our ``group" category includes both large clusters and moderate groups, which cover a wide range in HI deficiencies (Odekon et al. 2016), we do not find these larger differences.  (Selecting out only the large Coma cluster from our sample, on the other hand, we find a larger difference in average HI deficiency of  0.25 compared with void galaxies at about the same distance (90-110 Mpc). The average differences between galaxies in \textit{moderate}-density and low-density regions is much more subtle.  We also emphasize that, considering galaxies as a whole, most of the dependence of galaxy properties on environment can be attributed to differences in the \textit{ratio} of red and blue population, not on difference within either population separately (e.g. Balogh 2004). Here we consider changes within the blue/late-type population only. 

What are the physical implications of the trends we present here?  For example, could these differences in HI content imply a mechanism for galaxies to shift from the blue star-forming main sequence to the red population?  A difference in HI deficiency of 0.10 corresponds to a modest fractional difference of 20\% in HI mass, but these are average differences for populations with large variations.  It might be that the small difference in average gas content corresponds to a higher likelihood for galaxies in groups or filaments to reach a point where not enough gas is available to sustain star formation -- perhaps a threshold effect that would be discernible with deeper HI observations.

Our results do not directly distinguish between recent environmental effects (``nurture") and historical effects based on early conditions (``nature").  With only information about total HI content, we conclude that the specific physical mechanisms must somehow relate to large-scale structure in addition to local density.  Higher-resolution images might show signatures of specific, recent interactions with other galaxies or with the IGM.

Kuutma et al. (2017) focus on the dependence of the elliptical-to-spiral ratio on distance from the filament spine.  They find that, even at fixed local density, this ratio has a residual dependence on the distance to the filament (with more ellipticals close to the filament spine), particularly for high-mass galaxies.  They also find a shift in g-i color of about 0.05 \textit{within} the spiral population but (unlike Alpaslan 2016 and others), they do not see a dependence on stellar mass once local density is taken into account.  As shown in Figure 10,  we do find a dependence on stellar mass at small distances from the filament spine, and also a small shift in g-i color.  We suspect that the conflicting results depend on the specific process for correcting for mass and color simultaneously: because more massive galaxies tend to be redder, the overall dependence on distance from the filament spine could be attributed to either; here we choose the best model according to the BIC criterion.  We also note that these trends are strongest close to the filament spine; when we remove all group galaxies, regions closer than 1 Mpc from the filament spine are not adequately sampled, and we cannot see the mass dependence either.

The results presented in this paper support an observational picture of how galaxy properties depend on environment that has grown increasingly nuanced over the past few years.  In particular:

$\bullet$ Galaxies are ``pre-processed" in environments less extreme than that of clusters.  The quenching of star formation, as seen through reddening and HI deficiency,
apparently occurs in smaller groups and in filaments as well.

$\bullet$ While the most obvious differences in galaxy populations across different environments relate to the \textit{fraction} of red versus blue galaxies,  we can also detect trends in color and gas content with environment, \textit{within} the blue population.   

$\bullet$ While many properties of star-forming galaxies depend strongly on their mass, even at fixed stellar mass there is evidence for the quenching of star formation in dense environments. 

\acknowledgements

We thank an anonymous referee for comments that improved the paper. This work was funded in part by National Science Foundation Awards AST-1211005 and AST-1637339. MPH also acknowledges support from NSF/AST-1107390 and the Brinson Foundation. 

Funding for the Sloan Digital Sky Survey IV has been provided by
the Alfred P. Sloan Foundation, the U.S. Department of Energy Office of
Science, and the Participating Institutions. SDSS-IV acknowledges
support and resources from the Center for High-Performance Computing at
the University of Utah. The SDSS web site is www.sdss.org.

SDSS-IV is managed by the Astrophysical Research Consortium for the 
Participating Institutions of the SDSS Collaboration including the 
Brazilian Participation Group, the Carnegie Institution for Science, 
Carnegie Mellon University, the Chilean Participation Group, the French Participation Group, Harvard-Smithsonian Center for Astrophysics, 
Instituto de Astrof\'isica de Canarias, The Johns Hopkins University, 
Kavli Institute for the Physics and Mathematics of the Universe (IPMU) / 
University of Tokyo, Lawrence Berkeley National Laboratory, 
Leibniz Institut f\"ur Astrophysik Potsdam (AIP),  
Max-Planck-Institut f\"ur Astronomie (MPIA Heidelberg), 
Max-Planck-Institut f\"ur Astrophysik (MPA Garching), 
Max-Planck-Institut f\"ur Extraterrestrische Physik (MPE), 
National Astronomical Observatory of China, New Mexico State University, 
New York University, University of Notre Dame, 
Observat\'ario Nacional / MCTI, The Ohio State University, 
Pennsylvania State University, Shanghai Astronomical Observatory, 
United Kingdom Participation Group,
Universidad Nacional Aut\'onoma de M\'exico, University of Arizona, 
University of Colorado Boulder, University of Oxford, University of Portsmouth, 
University of Utah, University of Virginia, University of Washington, University of Wisconsin, 
Vanderbilt University, and Yale University.

\textit{Facilities:} \facility{Arecibo}, \facility{Sloan}

\clearpage


\begin{thebibliography}{}
	
\bibitem[Albareti(2016)]{alb16} Albareti et al. 2016, submitted to ApJS, arXiv:1608.02013
\bibitem[Alpaslan(2014)]{alp14} Alpaslan, M., Robotham, A., Driver, S. et al. 2014, \mnras, 438, 177
\bibitem[Alpaslan(2016)]{alp16} Alpaslan, M. Grootes, M., Marcum, P. M. et al. 2016, \mnras, 457, 2287
\bibitem[Aragon-Calvo(2016)]{ara16} Aragon-Calvo, M. A., Neyrinck, M. C., Silk, J. 2016, arXiv:1607.07881
\bibitem[Bahe(2013)]{bah13} Bahe, Y. M., McCarthy, I. G., Balogh, M. L., \& Font, A. S. 2013, \mnras, 430, 3017
\bibitem[Balogh(2004)]{bal04} Balogh, M., Baldry, I. K., Nichol, R., et al. 2004, \apj, 2, L101
\bibitem[Blanton(2005)]{bla05} Blanton, M. R., Eisenstein, D., Hogg, D. W., Schlegel, D. J. \& Brinkmann, J. 2005, \apj, 629, 143
\bibitem[Brough(2013)]{bro13} Brough, S., Croom, S., Sharp, R., et al. 2013, MNRAS, 435, 2903
\bibitem[Cen(2008)]{cen08} Cen, R., \& Riquelme, M. A. 2008, \apj, 674, 644
\bibitem[Chen(2017)]{che17} Chen, Y.-C., Ho, S., Mandelbaum, R. et al. 2017, \mnras, 466, 1880
\bibitem[Conselice(2013)]{con13} Conselice, C. J., Mortlock, A., Bluck, A. F. L., Grutzbauch, R., \& Duncan, K. 2013, \mnras, 430, 1051
\bibitem[Davies(1987)]{dav87} Davies, R.B. 1987, Biometrika, 74, 33
\bibitem[Dekel(2006)]{dek06} Dekel, A. \& Birnboim, Y. 2006, \mnras, 368, 2
\bibitem[Dressler(1980)]{dre80} Dressler, A. 1980, ApJ, 236, 351
\bibitem[Gavazzi(2013)]{gav13} Gavazzi,G., Savorgnan, G., Fossati, M., et al. 2013, A\&A, 553, 90
\bibitem[Giovanelli(1985)]{gio85} Giovanelli, R., \& Haynes, M. P. 1985, \apj, 292, 404
\bibitem[Giovanelli(1995)]{gio95} Giovanelli, R., Haynes, M. P.,  Salzer, J. J. et al. 1995,\aj, 110, 1059
\bibitem[Giovanelli(1997)]{gio97} Giovanelli, R., Haynes, M. P.,  Herter, T. et al. 1997, \aj 113, 53
\bibitem[Giovanelli(2005)]{gio05} Giovanelli, R., Haynes, M. P., Kent, B. R., Perillat, P. et al. 2005, \aj, 130, 2598 
\bibitem[Guo(2015)]{guo15} Guo Q., Tempel E., Libeskind N. I., 2015, ApJ, 800, 112
\bibitem[Hallenbeck(2012)]{hal12} Hallenbeck, G., Papastergis, E., Huang, S., et al. 2012, \aj, 144, 87 
\bibitem[Hallenbeck(2017)]{hal17} Hallenbeck, G., Koopmann, R., Giovanelli, R., Haynes, M. P., Huang, S., Leisman, L, \& Papastergis, E. 2017, \aj, 154, 58 
\bibitem[Haynes(1984)]{hay84} Haynes, M. P., \& Giovanelli, R. 1984, \aj, 89, 758
\bibitem[Haynes(2011)]{hay11} Haynes, M. P., Giovanelli, R., Martin, A. M., et al. 2011, AJ, 142,170
\bibitem[Helsel(2012)]{hel12} Helsel, D.R. 2012, Statistics for Censored Environmental Data using Minitab and R (2nd edition; Hoboken, NJ: Wiley)
\bibitem[Grutzbauch(2011)]{gru11} Grutzbauch, R., Chuter, R. W., Conselice, C. J., et al. 2011, MNRAS, 412, 2361
\bibitem[Jones(2016)]{jon16} Jones, M. G., Papastergis, E., Haynes, M. P., \& Giovanelli, R. 2016, \mnras, 457, 4393
\bibitem[Kleiner(2017)]{kle17} Kleiner, D., Pimbblet, K. A., Jones, D. H., Koribalski, B. S., \& Serra, P., 2017, \mnras, 466, 4692 
\bibitem[Kuutma(2017)]{kuu17} Kuutma, T., Tamm, A., Tempel, E., 2017, \aa, 600, L6		
\bibitem[Lai(2016)]{lai16} Lai, C.-C. Lin, L., Jian, H.-Y., et al. 2016, \apj, 825, 40
\bibitem[Laigle(2017)]{lai17} Laigle, C., Pichon, C., Arnouts, S. et al, 2017,\mnras, submitted
\bibitem[Lintott(2008)]{lin08} Lintott, C., Schawinski, K., Solsar, A., et al. 2008, \mnras, 389, 1179	
\bibitem[Moore(1995)]{moo95} Moore, B., Katz, N., Lake, G., Dressler, A., \& Oemler, A. 1996, Nature, 379, 613
\bibitem[Moorman(2016)]{moo16} Moorman, C.M., Moreno, J., White, A., Vogeley, M.S., Hoyle, F., Giovanelli, R. \& Haynes, M.P. 2016, ApJ 831, 118
\bibitem[Mouhcine(2007)]{mou07} Mouhcine, M., Baldry, I. K., \& Bamford, S. P. 2007, \mnras, 382, 801
\bibitem[Odekon(2016)]{ode16} Odekon, M. C., Koopmann, R. A., Haynes, M. P., et al. 2016, \apj, 824, 110 
\bibitem[Park(2007)]{par07} Park, C., Choi, Y.-Y., Vogeley, M. S., Gott, J. R., \& Blanton, M.R. 2007, 658, 898
\bibitem[Postman(1984)]{pos84} Postman, M., \& Geller, M. J. 1984, \apj, 281, 95 
\bibitem[Poudel(2017)]{pou17} Poudel, A., Heinämäki, P., Tempel, E., Einasto, E., Lietzen, H., Nurmi, P. 2017, A\&A, 597, A86
\bibitem[Raftery(1995)]{raf95} Raftery, A. E. 1995, Sociological Methodology, 25, 111
\bibitem[Rawle(2013)]{rqw13} Rawle, T.D., Lucey, J.R., Smith, R.J., \& Head, J.T.C.G. 2013, \mnras, 433, 2667
\bibitem[Saintonge(2007)]{san07} Saintonge, A. 2007, \aj, 133, 2087 
\bibitem[Sanchez(2014)]{san14} S´anchez Almeida, J., Elmegreen, B. G., Mu˜noz-Tu˜n´on, C., \&
Elmegreen, D. M., 2014, A\&A Rev., 22, 71
\bibitem[Schlafly(2011)]{sch11} Schlafly \& Finkbeiner, 2011, \apj, 737, 103
\bibitem[Schwarz(1978)]{sch78} Schwarz, G. E. 1978, \textit{Annals of Statistics}, 6 (2), 461
\bibitem[Shao(2007)]{sha07} Shao, Z., Xiao, Q., Shen, S., Mo, H. J.,Xia, X., \& Den, Z. 2007, \apj, 659, 1159
\bibitem[Solanes (2001)]{sol01} Solanes, J. M., Manrique, A., Garcia-Gomez, C., et al. 2001, \apj, 548, 97
\bibitem[Springob(2007)]{spr07} Springob, C. M., Masters, K. L., Haynes, M. P., Giovanelli, R., \& Marinoni, C. 2007, ApJS, 172, 599
\bibitem[Tanaka(2004)]{tan04} Tanaka, M., Goto, T., Okamura, S., Shimasaku, K., \& Brinkmann, J. 2004, \aj, 128, 2677
\bibitem[Taylor(2011)]{tay11} Taylor, E. N., Hopkins, A. M., Baldry, I. K., et al. 2011, \mnras, 418, 1587
\bibitem[Tempel(2015)]{tem15} Tempel E., Guo Q., Kipper R., Libeskind N. I., 2015, MNRAS, 450, 2727
\bibitem[Thilker(2004)]{thi04} Thilker, D. A., Braun, R., Walterbos, R. A. M., Corbelli, E., Lockman, F. J., Murphy, E., \& Maddelena, R. 2004, \apj, 601, L39 
\bibitem[Tonneson(2009)]{ton09} Tonneson, S., \& Bryan, G. L. 2009, \apj, 694, 789
\bibitem[Toribio(2011)]{tor11} Toribio, M. C., Solanes, J. M., Giovanelli, R., Haynes, M. P., \& Martin, A. M. 2011, \apj, 732, 93
\bibitem[Yang(2007)]{yan07} Yang, X., Mo, H. J., van den Bosch, F. C., Pasquali, A., Li, C., \& Barden, M. 2007, \apj, 671, 153
\bibitem[Zhang(2015)]{Zha15} Zhang Y., Yang X., Wang H., Wang L., Luo W., Mo H. J., van den Bosch, F. C., 2015, ApJ, 798, 17
\end{thebibliography}
\end{document}